\documentclass[sigplan,nonacm]{acmart}
\usepackage{xspace}
\usepackage{subcaption}
\usepackage{enumitem}
\usepackage{microtype}
\newcommand{\SysName}{Robion\xspace}
\newcommand{\SLOAT}{SLO-AT\xspace}
\newcommand{\SMR}{SLO-MR\xspace}

\newcommand{\Omnihigherload}{6.7}
\newcommand{\Monhigherload}{1.5}
\newcommand{\Monmultismr}{2.1}
\newcommand{\Largescalerobots}{64\xspace}

\begin{document}

\title{Efficient Vision-Language-Action Management and Serving for Robot Factories}

\author{Dionysios Adamopoulos}
\affiliation{%
  \institution{Max Planck Institute for Software Systems}
    \country{}
}

\author{Nattapol Chanpaisit}
\affiliation{%
  \institution{Max Planck Institute for Software Systems}
    \country{}
}

\author{Basel Fakhri}
\affiliation{%
  \institution{Max Planck Institute for Software Systems}
    \country{}
}

\author{Christina Giannoula}
\affiliation{%
  \institution{Max Planck Institute for Software Systems}
  \country{}
}

\begin{abstract}

Vision-Language-Action (VLA) models show impressive robotic manipulation capabilities via a two-stage design: a Vision-Language Model (VLM) stage followed by an Action Diffusion Transformer (ADiT) stage. Since robots must meet strict latency deadlines and Service-Level Objectives (SLOs) for safety and task effectiveness, VLA inference is inherently latency-critical. Meeting these SLOs requires high-end GPUs, yet weight, cost, and power constraints preclude integrating such GPUs on-robot. Recent works therefore offload VLA inference to local edge servers that serve many robots across a VLA model. However, current VLA systems lack support for multi-request, multi-model execution on a multi-GPU server under SLOs, while existing serving systems for multi-stage models are optimized for throughput and stage disaggregation across separate GPUs, which are ill-suited for the millisecond-scale stages of VLA models.
We design \SysName, the first VLA serving and management system for multi-robot, multi-model requests on multi-GPU edge servers that meets SLOs. \SysName serving engine disaggregates the VLM and ADiT stages within a GPU via two streams, dynamically restricting the SMs on VLM stream so ADiT always finds SMs to run alongside it, and co-locates multiple models by sharing these streams across them, prioritizing requests by least remaining SLO time. \SysName's management engine enables flexible model placements on multi-GPU servers, and integrates an intelligent traffic controller that maximizes per-model batching under the chosen placement while bounding each GPU's load to meet SLOs.
For individual models, \SysName serves on average \Omnihigherload$\times$ and \Monhigherload$\times$ higher robot load within 98\% SLO attainment over vLLM-Omni, the most widely used multi-stage serving system, and Monolithic, which runs VLM and ADiT as a single pipeline, respectively.
In a large-scale experiment of serving 8 different models on a 4-GPU server, \SysName can serve up to \Largescalerobots robots within 98\% SLO attainment.

\end{abstract}

\maketitle

\section{Introduction}

Physical Artificial Intelligence (AI) has emerged as a new frontier of AI, enabling autonomous agents, such as robots, to perceive complex environments, reason about sophisticated tasks, and execute actions in real-world settings.
These capabilities powered by robotics AI models~\cite{11164279,Ma_2026,shao2025largevlmbasedvisionlanguageactionmodels,zhong2025surveyvisionlanguageactionmodelsaction,Black2025pi0,Black2025pi05,Nvidia2025GROOT,Shukor2025Smolvla,Cai2026XiaomiRobotics0,XiaomiRoboticsTeam2026XiaomiRobotics1,Jiang2026ROSA,dream0,alpamayo,fastwam} that integrate perception, reasoning, planning, and action.
As physical AI matures, highly automated robot AI factories~\cite{Jiang2026ROSA} are expected to become increasingly prevalent. Leading companies, including BMW~\cite{figure2026bmw}, Amazon~\cite{agility2023amazon}, and DHL ~\cite{bostondynamics2024stretchdhl}, are incorporating robotic fleets across manufacturing facilities and warehouses, where robots perform assembly-line tasks.

At the forefront of physical AI models are Vision-Language-Action (\textbf{VLA}) models~\cite{11164279,Ma_2026,shao2025largevlmbasedvisionlanguageactionmodels,zhong2025surveyvisionlanguageactionmodelsaction,Black2025pi0,Black2025pi05,Nvidia2025GROOT,Shukor2025Smolvla,Cai2026XiaomiRobotics0,XiaomiRoboticsTeam2026XiaomiRobotics1,Jiang2026ROSA}.
VLA models connect visual perception and language-based reasoning to physical action, translating observations and task instructions into manipulation actions. VLA models are unlocking increasingly sophisticated capabilities, including object relocation ~\cite{Jiang2026ROSA,Shukor2025Smolvla,Cai2026XiaomiRobotics0,Black2025pi0}, sorting ~\cite{Jiang2026ROSA,Cai2026XiaomiRobotics0}, packing ~\cite{Black2025pi05,Black2025pi0}, and assembling ~\cite{Nvidia2025GROOT,Black2025pi0,Jiang2026ROSA}, establishing them as a critical foundation for robot AI factories.

Prior works~\cite{pohland2026offload,Jiang2026ROSA,Jiang2026HowFastCanIRunVLA}  propose deploying a local edge server within a robotic factory to run VLA inference for many robots, driven by three factors. 
First, VLA inference is inherently latency-critical, subject to strict latency and safety guarantees; missing latency deadlines can cause task failures, e.g., stop-and-go behavior or movement jerkiness~\cite{Bansal2026ActionChunkScheduling,guo2026actioncontrolnetlightweightdelayaware,Zhao2025VLARAIL,pohland2026offload}, unsafe navigation~\cite{hirose2026asyncvlaasynchronousvlafast}, or broader disruption, e.g., halting the assembly line~\cite{Jiang2026ROSA,yan2026agilevlafewshotindustrialpose}. 
Meeting these deadlines requires powerful GPUs: Fig.~\ref{fig:edge_vs_server} shows that edge GPUs, such as the Jetson AGX Thor, incur up to 3$\times$ higher latency than high-end GPUs (RTX 6000 Pro or H100). 
Second, integrating  high-end GPUs onto robots is impractical due to strict weight, cost, power, and thermal constraints~\cite{Jiang2026ROSA,pohland2026offload}; the added weight can compromise robot stability, and the added cost may be prohibitive. 
Third, on-robot inference substantially increases power consumption, reducing robot battery life and operational duration by up to 45\% ~\cite{pohland2026offload}, resulting in battery replacements and higher operating costs that compound across a multi-robot factory.  
These three factors make the case for a \emph{local edge server} in a robot factory that serves VLA inference for many robots  at a cost-effective, practical deployment.

To realize this edge server deployment, we analyze VLA models and their deployment scenario on server-class GPUs, and identify three design considerations (§\ref{sec:key-objectives}) for efficient VLA serving.
First, since robots must adhere to strict latency deadlines~\cite{Jiang2026HowFastCanIRunVLA,lu2026fasterrethinkingrealtimeflow,huang2026ticvla, wang2026realtime,yu2026surveyefficientvisionlanguageactionmodels} for both safety and task effectiveness, an effective serving system must meet per-request Service-Level Objectives (SLOs). 
Second, VLA architectures~\cite{11164279,Ma_2026,shao2025largevlmbasedvisionlanguageactionmodels,zhong2025surveyvisionlanguageactionmodelsaction,Black2025pi0,Black2025pi05,Nvidia2025GROOT,Shukor2025Smolvla,Cai2026XiaomiRobotics0,XiaomiRoboticsTeam2026XiaomiRobotics1,Jiang2026ROSA} comprise two distinct stages: a Vision-Language Model (\textbf{VLM}) followed by an Action Diffusion Transformer (\textbf{ADiT}). The VLM stage is \emph{compute-intensive}, dominated by high-arithmetic-intensity GEMM kernels, while the ADiT stage is \emph{memory-intensive}, dominated by low-arithmetic-intensity small matrix and element-wise vector kernels. These distinct profiles motivate \textbf{stage disaggregation} in VLA serving: separating the VLM and ADiT into two independently managed execution pipelines with \emph{stage-specific} runtime optimizations.  
Third, VLA models are small-scale~\cite{Jiang2026HowFastCanIRunVLA}, typically 1-5 billion parameters (e.g., 450M for SmolVLA, 4.7B for Xiaomi-Robotics-0), requiring $\sim$10GB of memory including inference-time tensors, far below the 80-120GB capacity of server-class GPUs. Therefore, a cost-effective VLA serving system can co-locate multiple VLA models per server-class GPU to exploit its available memory capacity.

Prior works~\cite{Jiang2026ROSA,dai2026kairos,Bansal2026ActionChunkScheduling} handle model placement of various robotic models (VLAs, World Action Models, and others)~\cite{Jiang2026ROSA}, reduce the actions robots take to complete tasks~\cite{dai2026kairos}, or reorder VLA requests to favor robots with fewer pending actions~\cite{Bansal2026ActionChunkScheduling}. 
None of them addresses concurrent execution and GPU sharing across multiple robot requests and VLA models under SLOs.
A few existing serving systems~\cite{Yin2026vllmomni,Jha2026Mstar,sglang_omni_2026} for multi-stage models disaggregate heterogeneous stages across \emph{separate} GPU devices.
As we demonstrate in §\ref{sec:limitation_existing_systems} and §\ref{sec:evaluation}, these systems are neither suitable nor efficient for VLA models. 
First, they are optimized for throughput, whereas VLA serving is inherently latency-critical.
Second, VLA stages exhibit low GPU utilization, e.g., for $\pi_0$ model on an RTX 6000 Pro at batch size 4, average tensor core utilization is 52\% and only 15\% for the VLM and ADiT stages, respectively. Placing the two stages on separate GPUs thus leaves most GPU compute resources underutilized.
Third, even when both stages run on a single GPU, these systems incur substantial request inspection overhead: to batch requests, the ADiT stage continuously inspects a buffer to check for newly arrived requests that completed the VLM stage, and merges them into its currently running batch. The inspection overhead is minimal for models with stages spanning hundreds of milliseconds, but becomes dominant for VLA models, whose stages execute in only a few milliseconds.


We propose \SysName, the first VLA serving and management system to handle multi-robot requests for multiple models on a multi-GPU edge server while meeting SLOs, realizing the paradigm of local edge servers in robot factories.

\SysName \textbf{serving engine} has three components. First, we disaggregate the VLM and ADiT stages within a GPU using two CUDA streams, executing them concurrently across independent request batches to exploit their complementary resource demands and improve GPU utilization. Second, we execute VLM and ADiT stages in synchronized steps (\emph{locksteps}) and leverage spatial Streaming Multiprocessor (SM) partitioning: we restrict the number of SMs the compute-intensive VLM stream can occupy, so ADiT always finds SMs available to run alongside it. We design a profiling-assisted method that \emph{dynamically} selects this SM restriction for the VLM stream at each lockstep to further improve performance. Third, we share a single VLM stream and a single ADiT stream across multiple co-located models on a GPU, so a VLM stage of one model can run concurrently with an ADiT stage of a different model, and prioritize requests with the least remaining SLO time to provide \emph{SLO-aware} serving.

\SysName \textbf{management engine} has two components. First, we enable programmer-configurable placement across GPUs: models can be co-located on a GPU, replicated across GPUs, or served by partitioning the server's GPUs into disjoint groups, each serving fine-tuned variants of a single VLA architecture. Second, we design an intelligent traffic controller for heterogeneous robot populations across co-located and replicated models, formulating traffic assignment as an integer program that concentrates each model's requests onto a single model copy on a GPU to increase per-model batching, while bounding each GPU's load to prevent SLO violations.

We evaluate \SysName across four VLA models, heterogeneous robot populations, varying SLO deadlines, diverse  model co-location and placement configurations, and two hardware platforms.
For individual models, \SysName significantly outperforms both vLLM-Omni, the most widely used serving system with stage disaggregation, and \emph{Monolithic}, which executes VLM and ADiT as a single pipeline, serving \Omnihigherload$\times$ and \Monhigherload$\times$ more robot load within 98\% SLO attainment, respectively.
When co-locating multiple VLA models on a GPU, spanning fine-tuned variants and different architectures, and serving up to 18 robots, \SysName achieves \Monmultismr$\times$ lower SLO miss ratio than \emph{Monolithic} on average, while vLLM-Omni remains fully saturated (100\% miss ratio).
In a large-scale experiment of serving 8 different models on a 4-GPU server, \SysName can serve up to \Largescalerobots robots within 98\% SLO attainment.

Overall, we make the following contributions: 
\begin{itemize}[topsep=0pt,leftmargin=16pt]
\item We investigate the design considerations for efficient VLA serving and propose \SysName, the first SLO-aware, multi-model VLA serving and management system.
\item We disaggregate  VLM and ADiT stages via GPU streams, running them concurrently for independent requests. We dynamically restrict the SMs of the VLM stage, so ADiT always finds SMs to run alongside it. We share a VLM and an ADiT stream across co-located models of a GPU, and prioritize requests with the least remaining SLO time.
\item We enable flexible model placement on GPUs, and design a traffic controller for heterogeneous robot populations across co-located and replicated models that increases per-model batching and bounds GPU loads to prevent SLO violations.
\item 
We evaluate \SysName on diverse models, robot populations, and deployments, showing significant SLO attainment improvements over existing serving systems.
\end{itemize}
\vspace{-4pt}
\section{Background \& Motivation}\label{sec:background-motivation} 
\vspace{-1pt}
\subsection{Vision-Language-Action (VLA) Models}\label{sec:vla-background}

VLA models ~\cite{11164279,Ma_2026,shao2025largevlmbasedvisionlanguageactionmodels,zhong2025surveyvisionlanguageactionmodelsaction,Black2025pi0,Black2025pi05,Nvidia2025GROOT,Shukor2025Smolvla,Cai2026XiaomiRobotics0,XiaomiRoboticsTeam2026XiaomiRobotics1,Jiang2026ROSA} have emerged as the leading approach for robotic manipulation tasks, such as object relocation ~\cite{Jiang2026ROSA,Shukor2025Smolvla,Cai2026XiaomiRobotics0,Black2025pi0}, sorting ~\cite{Jiang2026ROSA,Cai2026XiaomiRobotics0}, packing ~\cite{Black2025pi05,Black2025pi0}, and assembling ~\cite{Nvidia2025GROOT,Black2025pi0,Jiang2026ROSA}. A VLA model takes as input camera images and a text prompt, and outputs an action chunk for the robot to execute.

\begin{figure}[H]  
    \vspace{-3pt}
    \centering
    \includegraphics[width=\linewidth]{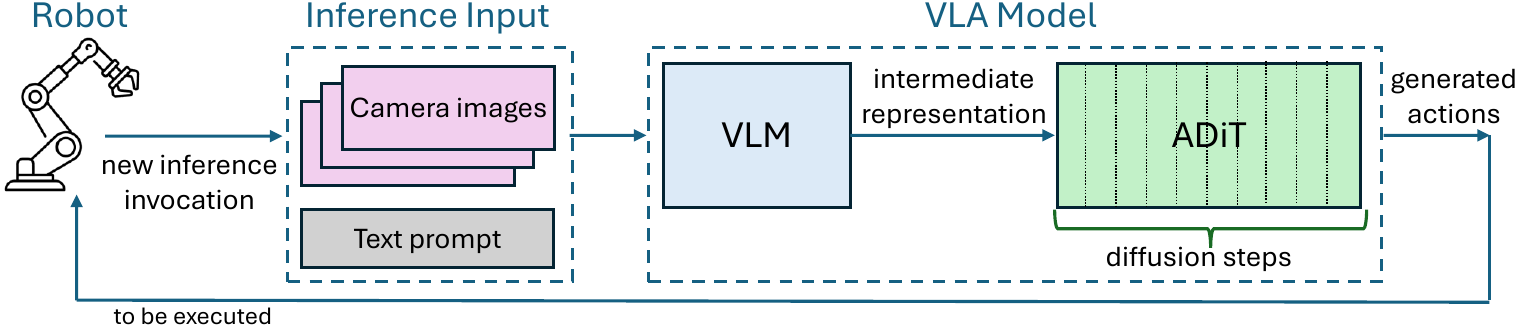}
    \vspace{-22pt}
    \caption{High-level architecture of a VLA model and its iterative inference loop with a robot.}
    \vspace{-10pt}
    \label{fig:vla_bckgrnd}
\end{figure}

State-of-the-art VLA models~\cite{Black2025pi0, Black2025pi05,Nvidia2025GROOT,Shukor2025Smolvla,Cai2026XiaomiRobotics0,XiaomiRoboticsTeam2026XiaomiRobotics1} typically follow a two-stage architecture of a Vision–Language Model (\textbf{VLM}) and an Action Diffusion Transformer (\textbf{ADiT}), as shown in Fig.~\ref{fig:vla_bckgrnd}. During inference, VLM processes the camera images and text prompt, and creates intermediate tensors (e.g., hidden states or a small KV cache), that are then forwarded to ADiT.  ADiT executes a fixed number of diffusion steps (typically 5-10) to produce a chunk of 10-50 future actions~\cite{Shukor2025Smolvla,Black2025pi0,Black2025pi05,Nvidia2025GROOT,Cai2026XiaomiRobotics0,XiaomiRoboticsTeam2026XiaomiRobotics1}. \textbf{Key characteristic} of VLA models is that both stages are lightweight, each typically takes a few milliseconds on a high-end GPU ~\cite{Black2025pi0,Jiang2026HowFastCanIRunVLA}, 
unlike LLMs that can take several seconds ~\cite{chittyvenkata2025moeinferencebenchperformanceevaluationmixture}.


Robotic manipulation proceeds in repeated rounds of VLA inference and action execution~\cite{Jiang2026HowFastCanIRunVLA,Black2025pi0,Shukor2025Smolvla,Jiang2026ROSA,dai2026kairos}: each round, the model processes the latest camera observations (the text prompt stays fixed for the task) and generates a new action chunk, which the robot executes sequentially before the next invocation. Since each call yields multiple future actions, the model need not run continuously: e.g., $\pi_0$ is invoked once every 500ms (2Hz), while inference runtime is $\leq$100ms~\cite{Black2025pi0}.

\subsection{\textbf{The Edge Server for VLA Inference }}\label{sec:local-server} 
\vspace{-1pt}

VLA inference is an inherently latency-critical workload~\cite{Jiang2026HowFastCanIRunVLA,lu2026fasterrethinkingrealtimeflow,huang2026ticvla, wang2026realtime,yu2026surveyefficientvisionlanguageactionmodels}, subject to strict safety deadlines and Service-Level Objectives (SLOs). Failing to meet these deadlines 
can cause task failures, stop-and-go behavior or movement jerkiness~\cite{Bansal2026ActionChunkScheduling,guo2026actioncontrolnetlightweightdelayaware,Zhao2025VLARAIL,pohland2026offload}, unsafe navigation~\cite{hirose2026asyncvlaasynchronousvlafast}, or broader environment disruption, 
e.g., halting assembly lines in robot factories~\cite{Jiang2026ROSA,yan2026agilevlafewshotindustrialpose}.

\begin{figure}[H]  
    \vspace{-4pt}
    \centering
    \includegraphics[width=\linewidth]{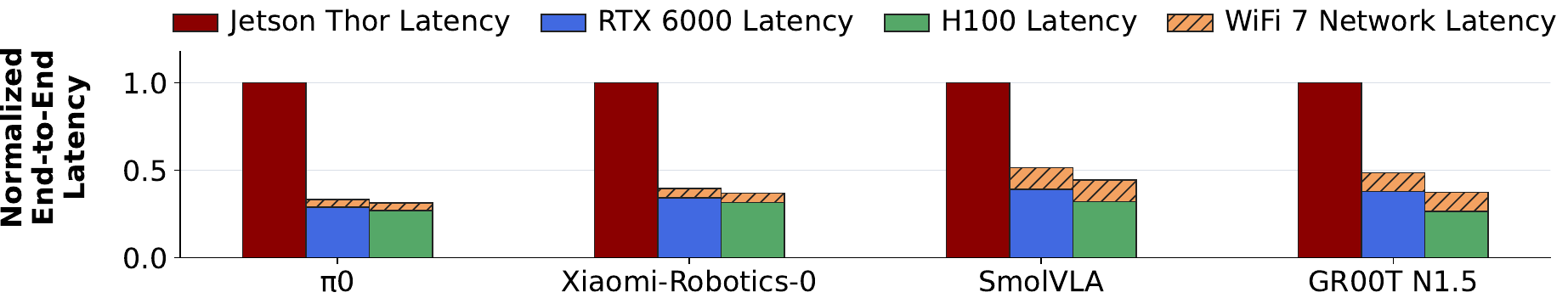}
    \vspace{-20pt}
    \caption{End-to-end inference latency for various VLA models across three different deployment scenarios.}
    \vspace{-8pt}
    \label{fig:edge_vs_server}
\end{figure}

Due to strict weight, cost, and power constraints ~\cite{Jiang2026ROSA,pohland2026offload}, robots typically cannot integrate high-end GPUs onboard, 
instead rely on low-area embedded GPUs, e.g., the NVIDIA Jetson series.
Prior works~\cite{pohland2026offload,Jiang2026ROSA,Jiang2026HowFastCanIRunVLA} show that offloading inference to edge servers with GPUs more powerful than what robots can integrate balances compute demand with VLA latency constraints. 
Fig.~\ref{fig:edge_vs_server} shows end-to-end inference latency for 
4 models on three deployment scenarios: (i) on-robot inference on a Jetson AGX Thor (70W power budget, no network cost); (ii) inference on an edge server with an RTX 6000 Pro GPU, communicating with robot over WiFi 7; and (iii) inference on  edge server with an H100 GPU, also over WiFi 7. 
Despite  network cost,
server inference on high-end GPU yields up to a $\sim$$3\times$ lower latency than on-robot inference, even for lightweight architectures, e.g., SmolVLA.

Running VLA inference on an embedded GPU per robot can also  be impractical due to power costs:  on-robot inference highly increases power consumption~\cite{pohland2026offload,Jiang2026ROSA}, reducing the robot battery life and operational duration by up to $\sim$$45\%$ ~\cite{pohland2026offload}. 
Instead, edge server offloading reduces inference latency, extends battery life,  improves cost efficiency, and allows a single server to serve multiple robots at high utilization, being a practical deployment target for VLA inference.

\vspace{-5pt}
\subsection{\textbf{Multi-Robot Serving Scenario}}\label{sec:scenarios}  
\vspace{-3pt}

A robot factory comprises a set of tasks (packing, assembling etc.), with $N_i$ robots assigned to task $i$. A robot sends inference requests to its task's VLA model $m_i$ at a task-dependent frequency $f_i$ ~\cite{Black2025pi0,Nvidia2025GROOT,Jiang2026ROSA}. 
Models of different tasks may be fine-tuned variants of the same VLA architecture~\cite{Black2025pi0, Nvidia2025GROOT,Cai2026XiaomiRobotics0,Shukor2025Smolvla} or different VLA architectures~\cite{Jiang2026ROSA}. 
A robot factory can  deploy a server (e.g., a local edge server) with one or more high-end GPUs, that hosts all models, while robots communicate with it over a low-latency wireless network, e.g., WiFi 7~\cite{Jiang2026ROSA,Jiang2026HowFastCanIRunVLA} to send inference requests and receive action chunks.

\vspace{-5pt}
\subsection{\textbf{Key Objectives for Efficient VLA Serving}}\label{sec:key-objectives} 
\vspace{-3pt}

Given the robot factory scenario, the VLA model characteristics, and the high-end GPU capabilities, we find three design considerations for  efficient VLA serving on high-end GPUs.

\noindent\textbf{1. SLO-Awareness:} Robots must adhere to strict latency deadlines, since increased latency reduces robot responsiveness and may cause task failure. Thus, an effective serving system must handle multi-robot requests while meeting per-request SLOs, e.g., achieving inference latency lower than that provided by an embedded GPU or on-robot SoC.

\noindent\textbf{2. Stage Disaggregation:}
VLA models have two architecturally distinct stages, VLM and ADiT. Fig.~\ref{fig:roofline-vllm}a shows roofline analysis of them on an RTX 6000 Pro GPU for 4 models.  VLM  is \textbf{compute-intensive}, dominated by high-arithmetic-intensity GEMM kernels, while ADiT is \textbf{memory-intensive}, dominated by small matrix and element-wise vector operations with low arithmetic intensity that underutilize tensor cores~\cite{Jiang2026ROSA}. Serving these stages as a \emph{single unified} inference pipeline can be inefficient: their distinct compute and memory profiles  call for separating them into independently managed execution pipelines (\emph{stage disaggregation)} to provide \emph{stage-specific} runtime optimizations. Moreover, executing the two stages \emph{independently} allows their execution to be overlapped across concurrent requests, exploiting their complementary resource demands to improve GPU utilization.


\noindent\textbf{3. Multi-Model GPU Co-Location:}
Unlike LLMs that have hundreds of billions of parameters, VLA models have only 1-5 billion parameters (e.g., 450M for SmolVLA, 4.7B for the larger Xiaomi-Robotics-0), requiring $\sim$10GB to store their weights and a few additional MBs for inference-time tensors. This footprint is far smaller than the memory capacity of high-end GPUs, e.g., 96GB on the RTX 6000 Pro, 80GB on the H100. Dedicating a high-end GPU to a single VLA model would leave most of its memory capacity unused. Effective VLA serving systems can co-locate multiple models on  GPU. 


\vspace{-5pt}
\subsection{\textbf{Limitations of Existing Systems}}  
\label{sec:limitation_existing_systems}
\vspace{-3pt}

ROSA~\cite{Jiang2026ROSA} handles the placement of heterogeneous robot models, including VLAs, World Action Models, safety models, and monitoring models, on multiple GPUs in a local cluster. Kairos~\cite{dai2026kairos} minimizes task completion time by modifying the VLA model itself to complete its task with fewer actions, while preserving task accuracy. Armory~\cite{Bansal2026ActionChunkScheduling} scheduler prioritizes requests from robots with few pending actions, and ensures that all robots in a factory continue to make progress at each timestep. 
However, none of these systems optimize VLA inference time to meet request-level SLOs or leverage high-end GPUs to serve multiple models concurrently, and all treat the two VLA stages as a single execution pipeline.

\begin{figure}[H]  
    \centering
    \includegraphics[width=\linewidth]{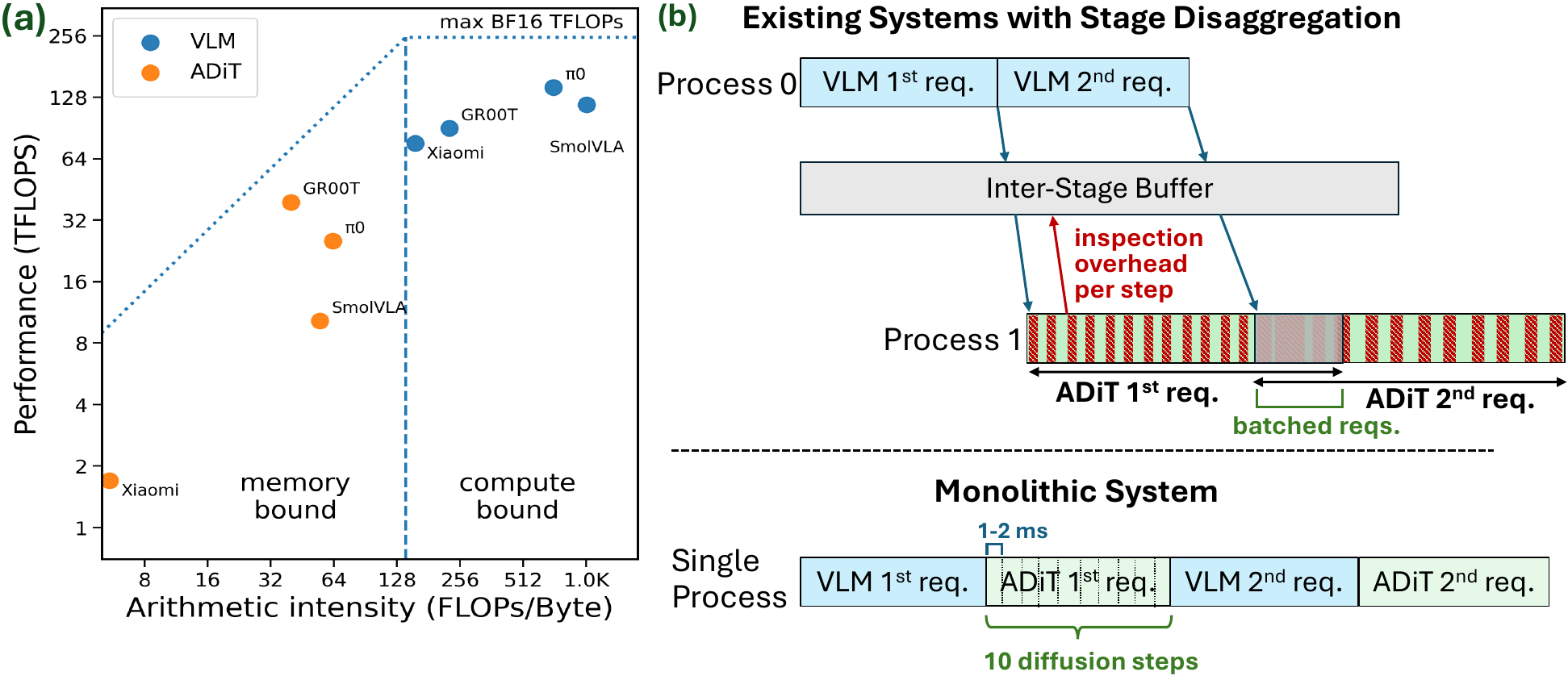}
    \vspace{-21pt}
    \caption{(a) Roofline analysis of VLM and ADiT stages across four VLA models on an RTX 6000 Pro GPU. (b) Existing systems with process stage disaggregation (top)  vs monolithic single-process execution (bottom).}
    \label{fig:roofline-vllm}
\end{figure}

Prior works~\cite{Yin2026vllmomni,Jha2026Mstar,sglang_omni_2026} design serving systems for models with  multiple distinct stages, e.g., image generation models~\cite{Xu2025Qwen-Omni,Deng2025BAGEL,zai2026glmimage} have an autoregressive language-model stage followed by a Diffusion Transformer (DiT) stage. 
These systems share a common design shown in Fig.~\ref{fig:roofline-vllm}b (top): they execute each stage as \emph{a separate process}, enabling independent scheduling strategies per stage, and use inter-stage buffers to transfer information (e.g., hidden states, KV caches) between stages. This  design  targets executing disaggregated stages on \emph{separate GPUs}, optimizes for throughput rather than latency, and is tailored to large-scale models whose stages run for hundreds of milliseconds, an order of magnitude longer than the millisecond-scale stages of VLA inference.


\vspace{-8pt}

These serving systems~\cite{Yin2026vllmomni,Jha2026Mstar,sglang_omni_2026} are not suitable nor efficient for VLA models, for two reasons. 
First, VLA model stages exhibit low GPU utilization, making it wasteful to run them on separate GPUs: profiling $\pi_0$ model (3B params.) on RTX 6000 Pro at batch size 4 shows 52\% average tensor core utilization for VLM stage, and only 15\% for ADiT stage (27\% at a fairly large batch size of 16), leaving most compute resources underutilized, especially in ADiT stage.
Second, even when both VLA stages are configured to run on one GPU, inter-stage communication costs remain significant: the throughput-oriented design of these systems requires the ADiT process 1 to inspect the inter-stage buffer at every diffusion step to batch newly arrived requests from the VLM process 0. This inspection cost (Fig.~\ref{fig:roofline-vllm}b (top)) is on the order of milliseconds, which is very costly for VLA models, whose diffusion steps are also on the order of milliseconds (1-2 ms each step, $\sim$10 steps total), leaving little room to amortize it. For example, in vLLM-Omni~\cite{Yin2026vllmomni,vllm_omni}, the most widely used serving system with stage disaggregation, this buffer inspection cost accounts for 30-35\% of total VLA inference latency. Consequently, even the monolithic approach \emph{without} stage disaggregation (Fig.~\ref{fig:roofline-vllm}b (bottom)) outperforms vLLM-Omni (See also §\ref{sec:eval-singlemodel}). Overall, existing serving systems with stage disaggregation are designed to provide throughput on large-scale compute-heavy multi-stage models rather than low latency on short, lightweight   VLA stages.

\section{\SysName: Design Overview}\label{sec:overview} 

\SysName is the first serving and management system to meet SLOs with high-end GPUs for multi-robot requests on multiple VLA models, enabling edge servers  in robot factories. 

\SysName serving engine integrates three  techniques:\\
\noindent\textbf{1. Fixed-Batch Intra-GPU Stage Disaggregation.} We disaggregate  VLM and ADiT stages within GPU as separate GPU streams to concurrently run them on independent requests. We support fixed-size batching, admitting requests only at the start of a stage, avoiding inspection and packing overheads during VLA stages' millisecond-scale execution.\\
\noindent\textbf{2. Guarded \& Dynamic SM Partitioning on Locksteps.} We run concurrent VLM and ADiT stages in \emph{locksteps}, exactly one VLM and one ADiT run per lockstep. We restrict the SMs the VLM stage can occupy, ensuring that ADiT  finds available SMs to run concurrently. We dynamically set this SM restriction based on each lockstep's VLM/ADiT batch-size combination to maximize performance.\\
\noindent\textbf{3. SLO-Aware Dual-Stream Multi-Model Serving.} We enable multi-model GPU co-location by sharing a VLM GPU stream and an ADiT GPU stream, among co-located models. We serve requests by prioritizing those with the least remaining time before their SLO expiration to meet SLOs.

\SysName management engine integrates two  techniques:\\
\noindent\textbf{1. Flexible Model Placement.}
We enable programmer-configurable model placement across the local server's GPUs. To efficiently accommodate different model architectures, we also support splitting GPUs into disjoint groups, each group serves only fine-tuned variants of a single VLA architecture.\\
\noindent\textbf{2. Adaptive Traffic Controller.} We propose a traffic adapter for varying robot populations sending requests to models co-located and/or replicated across GPUs. We formulate this multi-robot, multi-model, multi-GPU traffic assignment as integer program that increases per-model batching, while limiting each GPU's load to avoid SLO violations.

\section{\SysName: Detailed Design}\label{sec:mechanism} 

\subsection{\SysName Serving Engine}

\subsubsection{Fixed-Batch Intra-GPU Stage Disaggregation}\label{sec:mechanism-1}

For a VLA model, \SysName disaggregates VLM and ADiT stages within GPU to enable separate optimizations for each, overlap their executions for independent requests,  and improve GPU utilization. 
We use two asynchronous GPU streams, e.g., CUDA streams, one for VLM and one for ADiT.


However, supporting request batching introduces challenges.
First, continuous batching~\cite{orca,Kwon2023vllm} of throughput-oriented serving systems~\cite{orca,Kwon2023vllm,Yin2026vllmomni,holmes2024deepspeedfastgenhighthroughputtextgeneration,su2025seesaw, Agrawal2024TamingThroughputLatency} admits new requests into a running batch as they arrive. This requires inspecting for batching opportunities at each diffusion ADiT step  (§\ref{sec:limitation_existing_systems}), and this inspection cost can be disproportionately expensive relative to ADiT's step execution time, only 1-2 milliseconds.
Moreover, admitting new requests into a running ADiT batch requires packing their intermediate tensors, e.g., VLM-produced KV caches, with those of in-flight requests, incurring additional costs.
Third, strict SLOs of VLA inference constrain how large the batch size can be:  larger batches increase per-request latency. Overall, ADiT's short execution time and the latency-critical VLA nature  leave little room to amortize continuous batching costs.


Therefore, \SysName supports fixed-size stage batching for VLA stages: each stage executes on its own stream with a batch size \emph{fixed} for the duration of the stage execution, while the two streams may have different batch sizes during runtime, allowing VLM and ADiT executions to overlap across independent, variable-sized request batches. \SysName's fixed-size stage batching eliminates the per-ADiT-step inspection overheads that prior serving  systems~\cite{Yin2026vllmomni,Jha2026Mstar,sglang_omni_2026}  incur. 


\SysName uses CUDA graphs ~\cite{cudagraph}, recorded at a one-time initialization phase (§\ref{sec:mechanism-init}) to eliminate per-kernel CPU launch overheads, when repeatedly executing VLA stages.   ML serving systems ~\cite{Yin2026vllmomni,sglang_omni_2026,Kwon2023vllm,Jha2026Mstar,holmes2024deepspeedfastgenhighthroughputtextgeneration,su2025seesaw,Agrawal2024TamingThroughputLatency} also have an initialization phase for runtime optimizations, including using CUDA graphs. Since VLA inference is latency-critical, small batch sizes (e.g., 1-8) are preferable, which, in practice, makes \SysName record a small set of CUDA graphs during initialization, one for each combination of a stage and a batch size, and at runtime replays the corresponding graph on the appropriate stream.


\subsubsection{Guarded \& Dynamic  Spatial SM Partitioning on Locksteps}\label{sec:mechanism-2}

\SysName's stage disaggregation on two GPU streams makes VLM stage \emph{eligible} for concurrent execution with  ADiT for independent multi-request batches, but does not \emph{guarantee} it.
\underline{\emph{The Problem:}} The VLM stage includes large kernels with numerous thread blocks, while the ADiT stage small kernels with only a few thread blocks. The GPU thread block scheduler dispatches thread blocks based on residual per-SM capacity: once thread blocks of a running kernel fully occupy all SMs, no capacity remains~\cite{Tan2023GPUPool,cuda_guide} for a concurrently issued kernel to dispatch its thread blocks, thus their execution is serialized. In VLA models, a large kernel of the VLM stage  has enough thread blocks to saturate all SMs, blocking any concurrently issued kernel of the ADiT stage from dispatching its thread blocks until the VLM kernel releases SMs. \underline{\emph{The Opportunity:}} Conversely, a kernel of  ADiT stage has only a few thread blocks and occupies a few SMs, leaving the rest free for a concurrently issued kernel of VLM stage to dispatch its own thread blocks and run concurrently.


Our \textbf{Key Idea} is \emph{guarded SM partitioning} for the VLM stream: we restrict, at the granularity of the GPU stream, the number of SMs that the VLM stream, and hence every kernel of the VLM stage, can occupy, guaranteeing that the remaining SMs are always available for the ADiT stage's concurrently issued kernels to dispatch their thread blocks. This \underline{guarantees} concurrent execution of VLM and ADiT stages, exploiting their complementary resource demands: while VLM kernels operate on its fixed number of SMs, concurrently issued ADiT kernels dispatch their thread blocks to the remaining SMs. This, however, raises a \textbf{Key Challenge}: how many SMs should the VLM stream be restricted to?

\begin{figure}[t]  
    \centering
    \includegraphics[width=\linewidth]{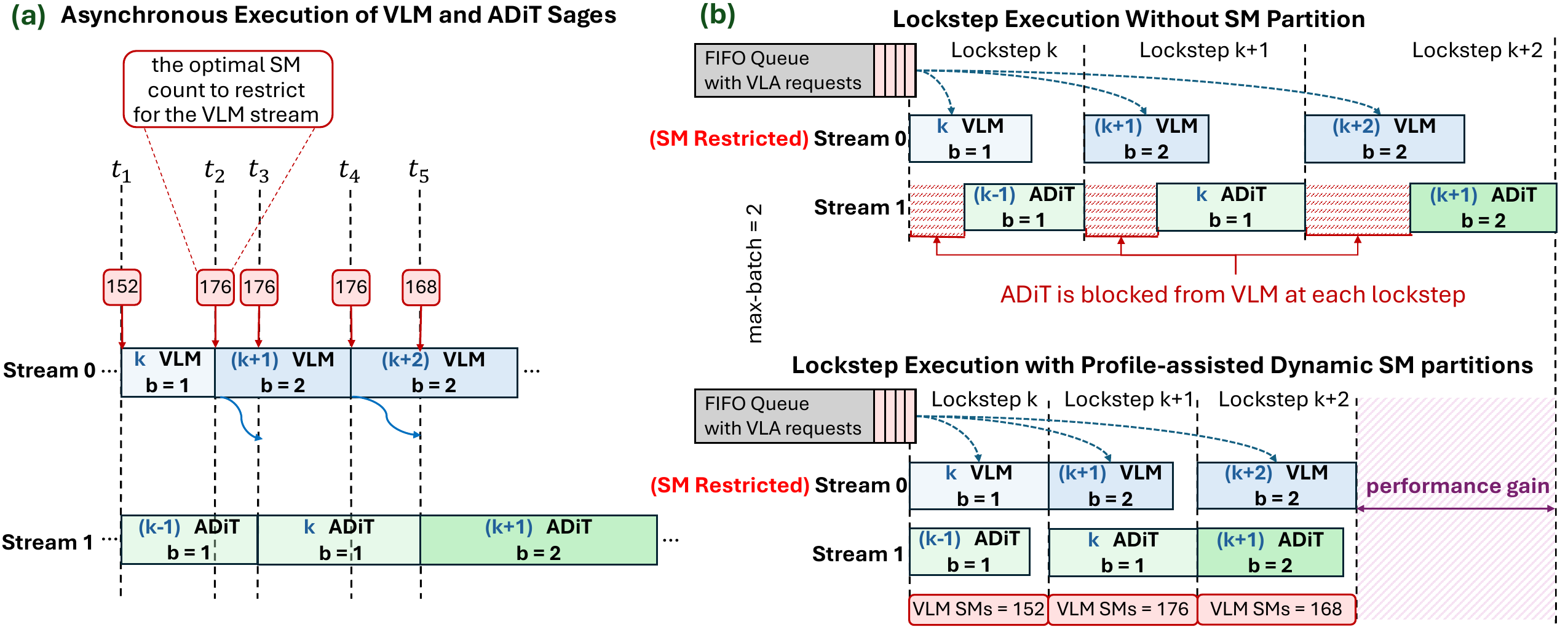}
    \vspace{-20pt}
    \caption{(a) An example asynchronous execution of VLM and ADiT stages with variable-sized multi-request batches. (b) Lockstep execution of VLM and ADiT stages without SM Partitioning versus lockstep execution of VLM and ADiT stages with profile-assisted dynamic SM partitioning.}
    \vspace{-14pt}
    \label{fig:mechanism_guarded}
\end{figure}

Fig.~\ref{fig:mechanism_guarded}a shows an example asynchronous execution of VLM and ADiT stages running on separate streams handling multi-request batches   assuming a GPU with 188 SMs. The red rectangles denote the \emph{optimal} number of SMs to which the VLM stream should be restricted at time $t_i$ to maximize performance, based on the batch sizes of the currently running VLM and ADiT stages.
At time $t_1$, both running VLM and ADiT stages have batch size 1,  and the optimal VLM restriction is 152 SMs. At time $t_2$, where the VLM and ADiT stages have batch size 2 and 1, respectively, the optimal VLM restriction is 176 SMs.
Dynamically setting the VLM SMs, e.g., at $t_2$, requires preempting the running VLM stage and relaunching it once the new SM restriction is reconfigured. This degrades performance in one of two ways: either having VLM executions without using CUDA graphs losing their performance benefits, since a stage recorded within a graph cannot be preempted, or if CUDA graphs are used,  preemption forces the whole stage to restart its execution from scratch, discarding any progress already made. In short,  dynamically setting the VLM SM restriction degrades performance, through either lost CUDA graph benefits or lost stage execution progress.


Our \textbf{Key Solution} (Fig ~\ref{fig:mechanism_guarded}b bottom) is the synergy of (i) lockstep stage execution with (ii) profiling-assisted dynamic SM restriction. We enable dynamic SM restriction on VLM stream without stage preemption, while leveraging CUDA graphs by executing VLM and ADiT stages as a sequence of \emph{locksteps}: at  lockstep $k$, the $k$-th VLM stage with batch size $b_k$, runs on one GPU stream concurrently with the ($k$-$1$)-th ADiT stage with batch size $b_{k-1}$ on the other stream. Incoming VLA requests are placed in a FIFO waiting queue.
At the start of lockstep $k$, we dequeue up to a \texttt{max\_batch} requests, and admit them as a fixed batch size $b_k$ to the VLM stage, while ADiT stage concurrently processes the ($k$-$1$)-th batch with batch size $b_{k-1}$.
Requests that complete the VLM stage during lockstep $k$-$1$ are held in a small temporary buffer and read by the ADiT stage at the start of lockstep $k$.
The \texttt{max\_batch} is set by the programmer, and bounds the number of requests batched per lockstep. If fewer than \texttt{max\_batch} requests are pending, we admit the largest power-of-2 batch size less than \texttt{max\_batch} to exploit pre-captured CUDA graphs. We use \texttt{max\_batch}=2 as the  default, that retains SLOs under high request traffic (Fig.~\ref{fig:batch-size-analysis}a).
We advance to the next lockstep once both stages finish. Since the VLM and ADiT batch sizes  vary across locksteps, the duration of each lockstep varies. Locksteps are implemented via two CPU threads, one per GPU stream, that launch the $k$-th VLM and ($k$-$1$)-th ADiT stages, respectively, and synchronize via a CPU thread barrier.

As shown in Fig.~\ref{fig:mechanism_guarded}b (top), within a lockstep, the larger VLM kernels occupy all SMs, blocking the smaller ADiT kernels, thus delaying them from being concurrently executed with VLM. However, each lockstep guarantees that \emph{exactly one} VLM and \emph{one} ADiT stage run concurrently within that time window. This allows us to restrict the VLM stream's SMs once, at the lockstep start, entirely avoiding mid-execution preemption of the running VLM stage. As shown in Fig.~\ref{fig:mechanism_guarded}b (bottom), restricting the VLM stream to occupy up to a fixed number of SMs allows  ADiT stage to begin execution immediately (at the lockstep start) on the rest SMs, rather than waiting for SMs to free up. Thus, the two stages make greater aggregated progress at each lockstep, shortening the lockstep's duration compared to that without SM restriction.


\SysName  finds the number of SMs to which the VLM stream is restricted at each lockstep, by performing offline profiling during its initialization phase (§~\ref{sec:mechanism-init}). Since one batched VLM stage co-executes with one batched ADiT stage per lockstep, and batch sizes are typically powers of 2 up to \texttt{max\_batch}, there are $O(log($\texttt{max\_batch}$)^2)$  possible stage co-execution combinations.
During profiling, we records CUDA graphs for both stages across all power-of-2 batch sizes up to \texttt{max\_batch}, then enumerate every combination of batch sizes for VLM and ADiT that can co-occur in a lockstep. For each combination, we exhaustively search over SM restrictions\footnote{Our profiler also considers the case of having \emph{no} SM restriction for VLM.} for the VLM stream, at the granularity supported by the GPU (e.g., multiples of 8 SMs on NVIDIA GPUs), profiling each candidate SM count restriction on the device to identify the one that minimizes the lockstep's execution time, i.e., the time for both co-running stages to complete. The best restriction found for each combination is stored in a lookup table. At runtime, when that combination occurs at a lockstep start, \SysName directly applies the corresponding SM count restriction to the VLM stream. 
For example, in Fig.~\ref{fig:mechanism_guarded}b (bottom), profiling found that limiting the VLM stream to 168 SMs minimizes lockstep time when VLM and ADiT use batch size 2. \SysName  applies this restriction at runtime, e.g., at lockstep $k$+$2$.
We implement the VLM stream SM restriction using CUDA Green Contexts~\cite{nvidia_cuda_driver} (we can use resource-partitioned execution contexts \cite{amdcontexts} for AMD GPUs).

\vspace{-6pt}
\subsubsection{SLO-Aware Dual-Stream Multi-Model Serving}\label{sec:mechanism-3}

A high-end GPU of an edge server can co-locate multiple VLA models (§\ref{sec:key-objectives}) serving  requests from robots for different tasks.
\SysName serves multiple co-located  models on a GPU by retaining its dual-GPU-stream design: a single \emph{shared} GPU stream for the VLM stage and a \emph{single} shared GPU stream for the ADiT stage are reused across all models. Within each lockstep, the VLM stream executes exactly one batch of requests issued for a single VLA model, while the ADiT stream concurrently executes exactly one batch of requests issued for a VLA model, the same one or a different one.


\SysName maintains its dual-stream design as shared across models for two reasons. 
First, running exactly one compute-intensive VLM stage concurrently with exactly one memory-intensive ADiT stage at each lockstep exploits their complementary compute and memory demands, improving GPU utilization across both resource dimensions. Instead, running multiple VLM or multiple ADiT stages concurrently creates contention for the same resources.
Second, \SysName aims to be \emph{SLO-aware}: different VLA models may have different SLOs, as they handle different tasks (e.g., packing vs. assembly) with different real-time responsiveness needs. This requires requests with stricter SLOs to be prioritized over those with more relaxed SLOs. Restricting each lockstep to a single VLM and a single ADiT stage allows \SysName choose which pending request to dispatch at the start of each lockstep, enabling SLO-aware prioritization of requests with tighter deadlines.


\SysName has one FIFO queue per co-located model and applies an \emph{earliest-SLO-first} policy: at the start of each lockstep, it computes, for each queue, the remaining time of its head request until the SLO deadline, i.e., the SLO deadline minus the time the request has already waited since its arrival, and selects the FIFO queue whose head request has the least remaining time. \SysName then dequeues a batch of up to \texttt{max\_batch} requests from that queue and admits them to the VLM stage, while the ADiT stage concurrently processes the batch that completed the VLM stage in the previous lockstep.

To handle multiple co-located models, \SysName extends its offline profiling to enumerate stage co-execution combinations across different VLA model pairs. 
This yields $O(U^2 \cdot log($\texttt{max\_batch}$)^2)$ combinations, where $U$ is the number of \emph{distinct} VLA architectures co-located on the GPU. When co-located models are fine-tuned variants of the \emph{same} architecture, all such model pairs share identical lockstep runtimes for a given batch size combination, so \SysName reuses the same profiling results across them. $U$ is small in practice.

\subsection{\SysName Management Engine}\label{sec:mechanism-4}

\subsubsection{Flexible Model Placement}

\SysName supports flexible, programmer-configurable model placement across the available server's GPUs. Programmers can co-locate models of different architectures and fine-tuned variants of the same architecture 
on a GPU, and/or replicate models to multiple GPUs. 
\SysName  also  supports partitioning the server's GPUs into disjoint groups, each group is dedicated to serving fine-tuned variants of a single VLA model architecture, allowing co-location of these variants within a GPU.

\subsubsection{Adaptive Traffic Controller}\label{sec:controller} 

Robot factories comprise heterogeneous task populations \cite{Jiang2026ROSA,Bansal2026ActionChunkScheduling}, where different models are assigned different numbers of robots (§\ref{sec:scenarios}). At initialization, \SysName takes as input the robot population $N_i$ and the request frequency $f_i$ for each served model $m_i$, and computes for every model $m_i$ and each GPU  $j$ hosting it, the number of robots $r_{ij}$ to assign to that GPU.


Since the total robot population for model $m_i$ is $N_i$, the robots assigned across all GPUs $G$ for model $m_i$ must sum to $N_i$ (with $r_{ij}$=$0$ for a GPU $j$ \emph{not} hosting the model $m_i$):

\[
\sum_{1 \leq j \leq G}r_{ij}=N_i \tag{1}
\]

\begin{figure}[H]  
    \vspace{-6pt}
    \centering
    \includegraphics[width=\linewidth]{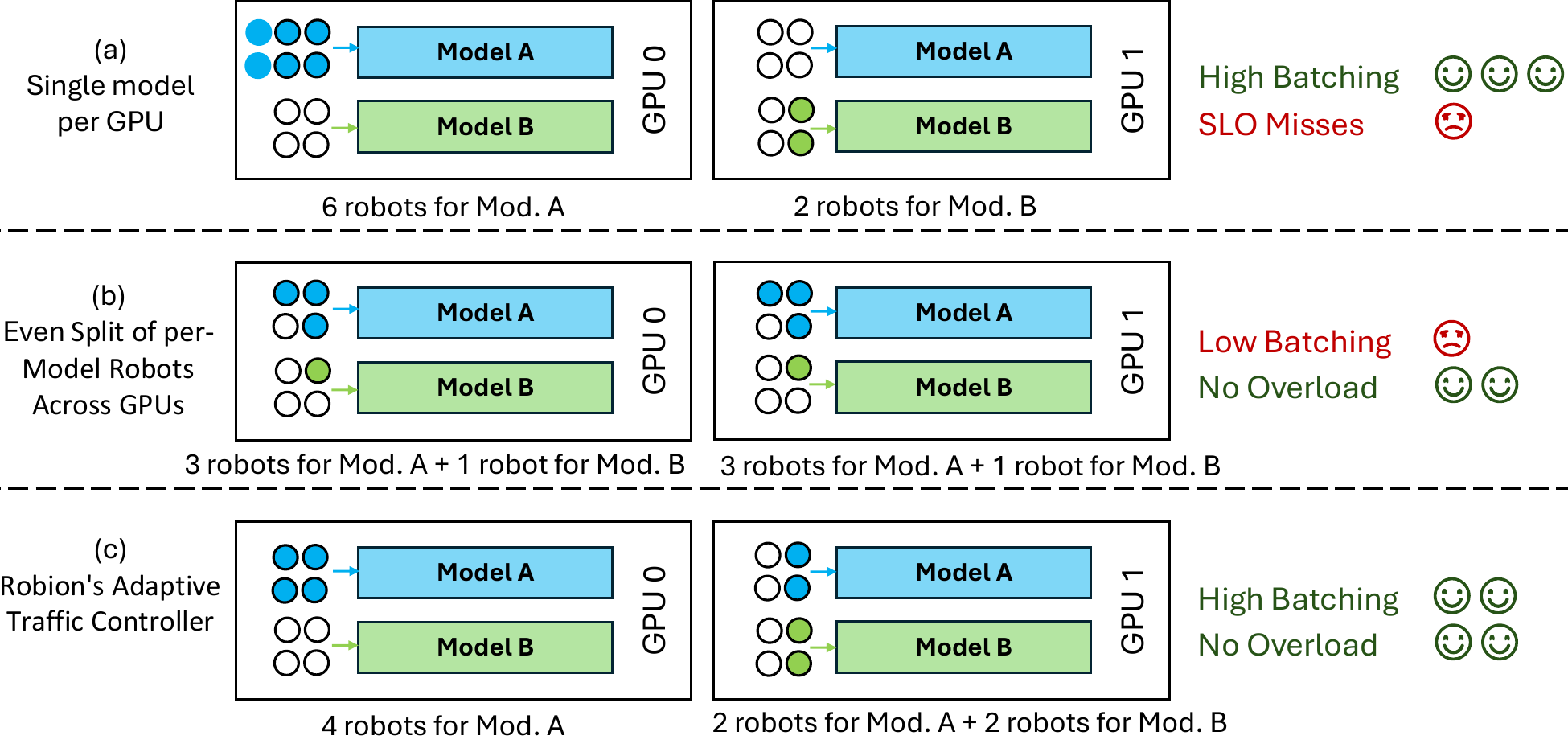}
    \vspace{-20pt}
    \caption{Traffic distribution strategies for two models on two GPUs: (a) each GPU serves only one model, (b) even robot split across GPUs, and (c) \SysName traffic adapter.}
    \vspace{-7pt}
    \label{fig:mechanism_traffic}
\end{figure}

Fig.~\ref{fig:mechanism_traffic} has a two-GPU server serving Models A and B of 6 and 2 robots, respectively. We assume each GPU can serve at most 4 robots, beyond this, queuing delays incur SLO misses.


Fig.~\ref{fig:mechanism_traffic}a shows a scheme where a GPU serves a single model: GPU 0 serves all 6 robots for Model A and GPU 1 serves all 2 robots for Model B. This scheme enables high  batching capabilities per model, since each model's FIFO queue accumulates sufficient pending requests from its dedicated traffic. However, GPU 0 serves 6 robots, exceeding its 4-robot capability, thus these robot requests incur SLO violations.


Fig.~\ref{fig:mechanism_traffic}b shows a scheme where each model's robots are split \emph{evenly} across the two GPUs: each GPU serves 3 robots for Model A and 1 robot for Model B. This allows neither GPU to exceed its 4-robot limit. However, splitting robot populations across GPUs reduces the per-model traffic each GPU receives: the queue for Model B at either GPU, backed by 1 robot, accumulates few requests, thus lower GPU utilization (smaller batch sizes used due to infrequent request arrivals).


\SysName integrates an intelligent traffic adapter, shown in Fig.~\ref{fig:mechanism_traffic}c, that combines the benefits of both prior schemes: high per-model batching and avoiding GPU overloads (keeping each GPU within its robot limit). It assigns each GPU to serve a single model whenever possible, concentrating that model's requests into one FIFO queue to enable high batching as in Fig.~\ref{fig:mechanism_traffic}a, while splitting a model's robots across additional GPUs as needed to stay within each GPU's limit, as in Fig.~\ref{fig:mechanism_traffic}b. In our example, \SysName assigns 4 robots for Model A to GPU 0, and the remaining 2 robots for Model A together with  2 robots for Model B to GPU 1. This concentrates traffic within per-model queues, e.g., GPU 0 is saturated with requests from 4 robots on Model A, achieving high batching while keeping both GPUs within their supported limits.


\noindent\textbf{Finding GPU Robot Limits.}\label{sec:limits}  \SysName needs to find the maximum number of robots $C_i$ a GPU can serve for each model $m_i$ while sustaining a target SLO attainment (e.g., 98\%). At initialization, \SysName benchmarks each model $m_i$ by simulating robot traffic on GPU as described in §\ref{sec:methodology}, and increasing the number of robots until finding the largest value that still meets the target SLO attainment. Benchmarking happens once per model and is parallelized on the server's GPUs. 


\noindent\textbf{Problem Formulation.} Let $M$ denote the total number of models served by \SysName. By capturing for a given model $m_i$ the maximum number of robots $C_i$ that a GPU $j$ can sustain, the ratio $\frac{r_{ij}}{C_i}$ shows how saturated a GPU is for traffic targeting the model $m_i$. To ensure that the assigned traffic to a GPU does not exceed its supported limit (avoiding SLO violations), we set the following constraint for every GPU $j$:

\[
\sum_{1 \leq i \leq M}\frac{r_{ij}}{C_i}\leq 1  \tag{2}
\]

To concentrate all received requests on a single queue, \SysName tries to minimize the sum of all pairwise products $(r_{ij} \cdot r_{kj})$ across all GPUs and distinct model pairs $(m_i, m_k)$: 

\[
min (\sum_{1 \leq j \leq G}\sum_{1 \leq i < k \leq M} r_{ij}\,r_{kj} )\tag{3}
\]

Each product $r_{ij}\cdot r_{kj}$ is non-zero only when robots targeting the distinct models $m_i$ and $m_k$ are sending requests to the same GPU $j$ (i.e., both variables $r_{ij}$ and $r_{kj}$ are non-zero). In contrast, if in each GPU $j$ only one term $r_{ij}$ is non-zero and others $0$, i.e., only robots targeting model $i$ send requests to GPU $j$, then every product term in (3) is zero (\emph{minimum}). Hence, this objective favors an assignment of robots where each GPU serves requests targeting a single VLA model.

Together, constraints (1) and (2), and the objective (3), define an integer programming problem, which \SysName solves during initialization. If constraints (1) and (2) cannot be satisfied, \SysName informs the programmer that they need to select a different model placement or add more GPUs to the server. We implement our solver using OR-Tools~\cite{ortools}.

\subsection{Initialization of \SysName}\label{sec:mechanism-init}

\SysName's one-time initialization comprises four parts: (a) \textit{graph capture} records GPU graphs for VLM and ADiT at each supported batch size (§\ref{sec:mechanism-1}), (b) \textit{profiling} finds and stores the VLM SM restriction that minimizes lockstep time for each combination of co-executed stages and batch sizes (§\ref{sec:mechanism-2}, §\ref{sec:mechanism-3}), (c) \textit{benchmarking} determines each model's maximum robot population that a GPU can serve under a target SLO attainment (§\ref{sec:controller}), and (d) \textit{solver execution} computes robot-to-GPU assignments using (c) robot limits and the configured model placement (§\ref{sec:controller}). For two VLA architectures with four fine-tuned variants each (8 models) on a 4$\times$RTX 6000 Pro server (two disjoint GPU groups, four variants co-located per GPU), parts (a)-(d) take $\sim$4 mins, $\sim$3 mins, $\sim$1 hour, and $\sim$10 s, respectively. Parts (b) and (c) rerun only when the model set, \texttt{max\_batch}, or GPU hardware changes, (d) reruns when robot populations change, and (a) reruns on every initialization, a short, standard overhead existing also in prior ML serving systems~\cite{Yin2026vllmomni,Jha2026Mstar,sglang_omni_2026}.

\vspace{-4pt}
\section{Evaluation}\label{sec:evaluation} 
\vspace{-2pt}
\subsection{\textbf{Evaluation Methodology}}\label{sec:methodology}

\noindent{\textbf{Models \& Datasets.}}
We evaluate four model architectures: $\pi_0$ (\textbf{PI0})~\cite{Black2025pi0}, SmolVLA \textbf{(SVLA)}~\cite{Shukor2025Smolvla}, GR00T N1.5 \textbf{(GR15)}~\cite{Nvidia2025GROOT}, and Xiaomi-Robotics-0 \textbf{(XR0)}~\cite{Cai2026XiaomiRobotics0}. Camera images and text instructions for requests are drawn from the LIBERO dataset~\cite{Liu2023LIBERO}.

\noindent{\textbf{Hardware Platforms.}}
We use two servers: one equipped with 4$\times$NVIDIA RTX 6000 Pro Blackwell GPUs (96GB each) and one with 4$\times$NVIDIA H100 GPUs (80GB each). Unless otherwise stated, we use the first server for our evaluations.

\noindent{\textbf{SLO Metrics.}}
We report SLO attainment (\textbf{\SLOAT}) /miss rates (\textbf{\SMR}). Unless otherwise stated, each model's SLO deadline is its inference latency on an NVIDIA Jetson AGX Thor GPU (70W power mode): 156, 117, 50, and 53 ms for PI0, XR0, SVLA, and GR15, respectively. To evaluate \SLOAT we compare each request's end-to-end latency, the sum of measured server-side inference latency and network latency, against the corresponding deadline. Unless otherwise stated, network latency follows the WiFi 7 configuration as in~\cite{Jiang2026HowFastCanIRunVLA}.

\noindent{\textbf{Robot Traffic.}} 
Robots targeting a VLA model send inference requests at frequency $f$. To simulate this traffic, we randomly sample an initial offset within time  $[0, 1/f)$ for each robot, after which it sends requests at frequency $f$. To account for variability in offset alignment, we repeat each experiment over five epochs, resampling robot offsets at the start of every epoch, and report the average results across epochs. Unless otherwise stated, we use $f$=2Hz.

\noindent{\textbf{Comparison Points.}}
We compare \SysName\ with \texttt{max\_batch}=$2$ over two systems: (i) \textbf{Monolithic} system executes the VLM and ADiT stages in a single, unified inference pipeline, and (ii) \textbf{vLLM-Omni} system~\cite{Yin2026vllmomni}, the most widely used serving system with model stage disaggregation, executes the VLM and ADiT stages as separate processes within a GPU.




\vspace{-2pt}
\subsection{\textbf{Single Model Serving on a GPU}}\label{sec:eval-singlemodel}
\begin{figure}[H]
    \centering
    \includegraphics[width=\linewidth]{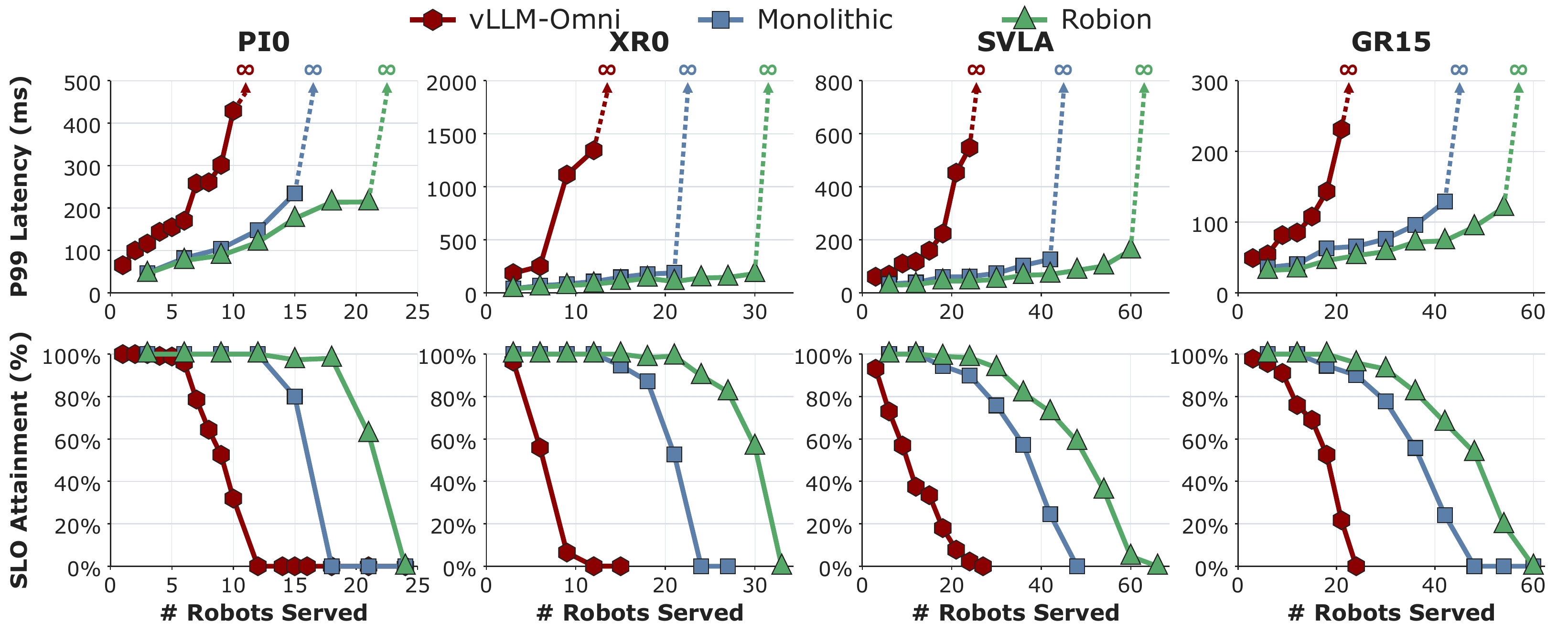}
    \vspace{-20pt}
    \caption{P99 latency (top) and \SLOAT (bottom) of all systems for various four models and robot loads. Dashed arrows show that the robot load has saturated the system.}
    \vspace{-9pt}
    \label{fig:single-model-single-gpu_rtx6000}
\end{figure}

\noindent\textbf{1) Performance.} Fig.~\ref{fig:single-model-single-gpu_rtx6000} compares the p99 latency and \SLOAT of all systems, when serving a single model on a single GPU under varying robot loads.

\vspace{-10pt}

We draw three findings. First, Monolithic outperforms vLLM-Omni across all models and robot loads, sustaining on average 4.4$\times$ more robot load within 98\% \SLOAT. This is because vLLM-Omni's per-diffusion-step inspection cost (§\ref{sec:limitation_existing_systems}) is expensive relative to VLA inference's millisecond-scale runtime, negating the benefits of stage disaggregation. 
Second, \SysName delivers the highest performance across all models and robot loads: at the highest robot load Monolithic sustains before saturation, \SysName achieves  1.4$\times$ lower p99 latency on average, and within 98\% \SLOAT, sustains on average \Omnihigherload$\times$ and \Monhigherload$\times$ more robot load than vLLM-Omni and Monolithic, respectively, across all models. 
Third, \SysName's efficiency enables a single high-end GPU to outperform on-robot inference on an embedded Jetson AGX Thor GPU, while serving multiple robots: for all models, \SysName achieves at least 98\% \SLOAT, while serving an average of 20 robots. This can offer large financial benefits: given the cost  of Jetson AGX Thor and RTX 6000 Pro is $\sim$5.5k USD and $\sim$16k USD, respectively, \SysName's serving capability translates to $\sim$6.8$\times$ cost reductions in hardware equipment. 
\underline{\textit{Overall}}, \SysName significantly outperforms existing serving systems, sustaining substantially higher robot loads, while consistently meeting SLOs across various  model architectures.

\begin{figure}[H]
    \centering
    \includegraphics[width=\linewidth]{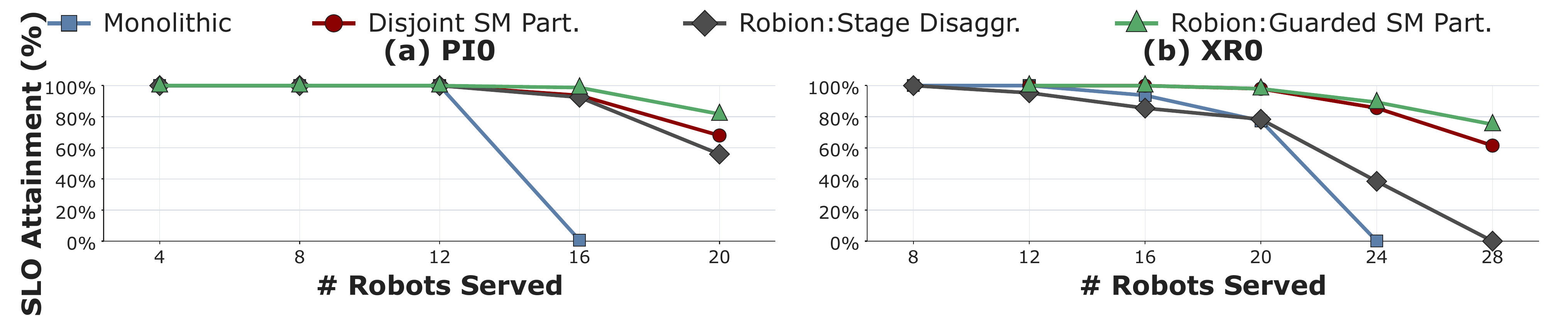}
    \vspace{-20pt}
    \caption{\SLOAT on PI0 and XR0 for Monolithic, disjoint SM partitioning, \SysName stage disaggregation only, and \SysName stage disaggregation $+$ guarded \& dynamic SM partitioning.}
    \vspace{-8pt}
    \label{fig:ablation-guarded-overlap}
\end{figure}

\noindent\textbf{2) Benefits of §\ref{sec:mechanism-1} and §~\ref{sec:mechanism-2} Key Techniques.} Fig.~\ref{fig:ablation-guarded-overlap} shows \SLOAT on PI0 and XR0 comparing (i) \emph{Monolithic}, (ii) \emph{Disjoint SM Part.} which uses \SysName's lockstep execution to dynamically partition SMs (via profiling) into two \textbf{disjoint}, non-overlapping sets for VLM and ADiT, (iii) \emph{\SysName Stage Disaggr.} which applies only intra-GPU stage disaggregation (§\ref{sec:mechanism-1}) and (iv) \emph{\SysName Guarded SM Part.} which applies both stage disaggregation (§\ref{sec:mechanism-1}) and VLM guarded \& dynamic SM partitioning on locksteps (§\ref{sec:mechanism-2}).

We make two observations. First, while \emph{\SysName Stage Disaggr.} increases \SLOAT over Monolithic, achieving 56\% at 20 robots for PI0 and 39\% at 24 robots for XR0, compared to 0\% for Monolithic for both models, \emph{\SysName Guarded SM Part.} improves even further, reaching 82\% and 89\% \SLOAT at the same loads and models. Second, \emph{Disjoint SM Part.} performs worse than \emph{\SysName Guarded SM Part.}, achieving 67\% at 20 robots for PI0 and 85\% at 24 robots for XR0. This is primarily because partitioning the RTX 6000 Pro's 188 SMs into two disjoint sets requires both sets to be multiples of 8, but 188 is not divisible by 8. Thus, \emph{Disjoint SM Part.} rounds down to 184 SMs (nearest multiple of 8), leaving 4 SMs unused.

\begin{figure}[H]
    \centering
    \includegraphics[width=\linewidth]{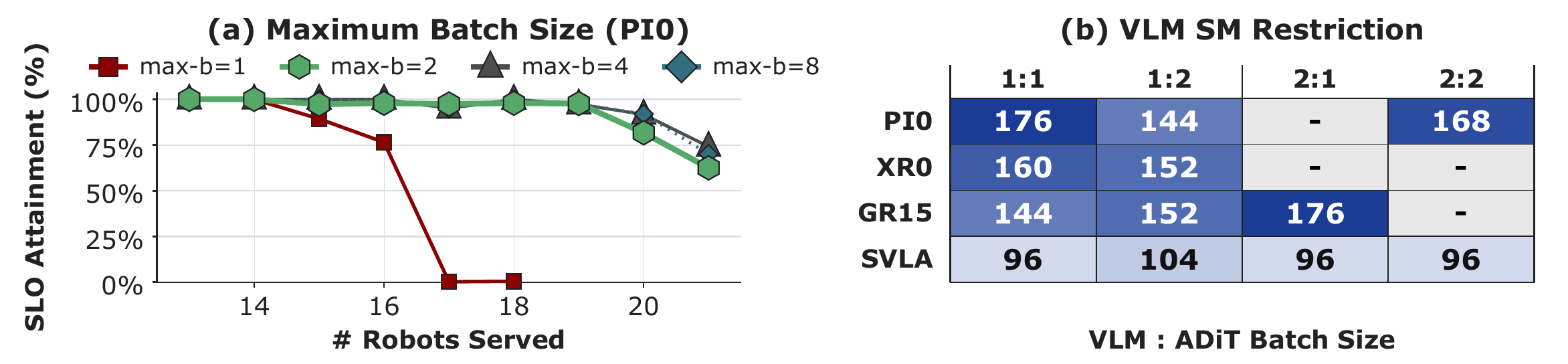}
    \vspace{-20pt}
    \caption{(a) \SysName's \SLOAT for PI0 across varying robot loads and \texttt{max\_batch} values. (b) \SysName's per-model selected VLM SM restrictions with \texttt{max\_batch}=2 for all batch size combinations for VLM and ADiT stages co-execution.}
    \label{fig:batch-size-analysis}
\end{figure}

\noindent\textbf{3) \texttt{Max\_Batch} Size and SM Restriction Analysis.}
Fig.~\ref{fig:batch-size-analysis}a shows \SysName's \SLOAT for PI0 across varying robot loads and \texttt{max\_batch} configurations. Setting \texttt{max\_batch=1} results in poor \SLOAT, when serving more than 14 robots, whereas \texttt{max\_batch=2} retains high \SLOAT even under high robot load (e.g., 98\% at 18 robots). 
Fig.~\ref{fig:batch-size-analysis}b shows the per-model number of SMs to which the VLM stage is restricted via \SysName's one-time profiling, with \texttt{max\_batch=2}, across all batch size combinations for VLM and ADiT stage co-execution. \emph{Dashes} denote cases in which \SysName does not restrict the SMs on VLM stage, as our profiling shows that the lockstep time is minimized without any SM restriction. We observe that different models and batch size combinations for stage co-execution require different SM count restrictions. Moreover, statically restricting the VLM stage to, e.g., 96 of the 188 available SMs reduces the robot load that can be served within 98\% \SLOAT by $\sim$18\% on average across all models, which highlights the necessity of \SysName's dynamic SM restriction.


\vspace{-4pt}

\subsection{\textbf{Multi-Model Co-Location and Serving on a GPU}}
\begin{figure}[H]
    \centering
    \includegraphics[width=\linewidth]{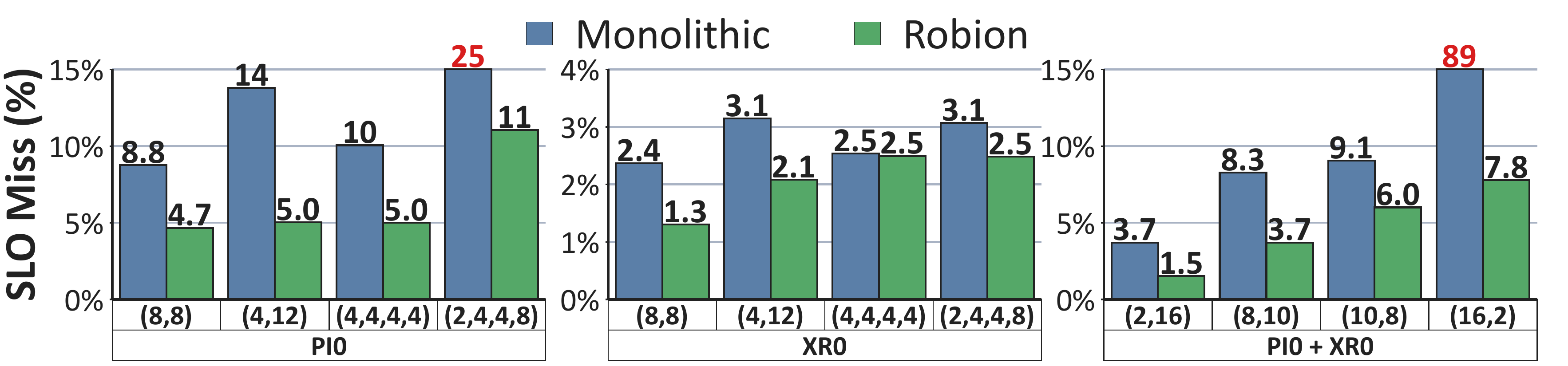}
    \vspace{-20pt}
    \caption{\SMR of Monolithic and \SysName for various configurations of co-located models on a single GPU and using different robot populations for each co-located model.}
    \vspace{-10pt}
    \label{fig:multiple-models-architectures}
\end{figure}

\noindent\textbf{1) Performance.} Fig.~\ref{fig:multiple-models-architectures} evaluates \SMR for Monolithic and \SysName when co-locating multiple models on a single GPU, including fine-tuned variants of the same architecture and of different architectures, with robot populations across co-located models either balanced or highly skewed. 
Each x-axis tuple denotes a configuration: its length gives the number of co-located models, and each entry gives that model's robot population. We omit vLLM-Omni, which is already saturated (100\% \SMR) across these configurations. \SysName outperforms Monolithic across all multi-model configurations, achieving on-average \Monmultismr$\times$ lower \SMR under various robot-population distributions. We conclude that \SysName effectively co-locates multiple models on a single high-end GPU, improving GPU utilization and cost-effectiveness.

\begin{figure}[H]
    \vspace{-1pt}
    \centering
    \includegraphics[width=\linewidth]{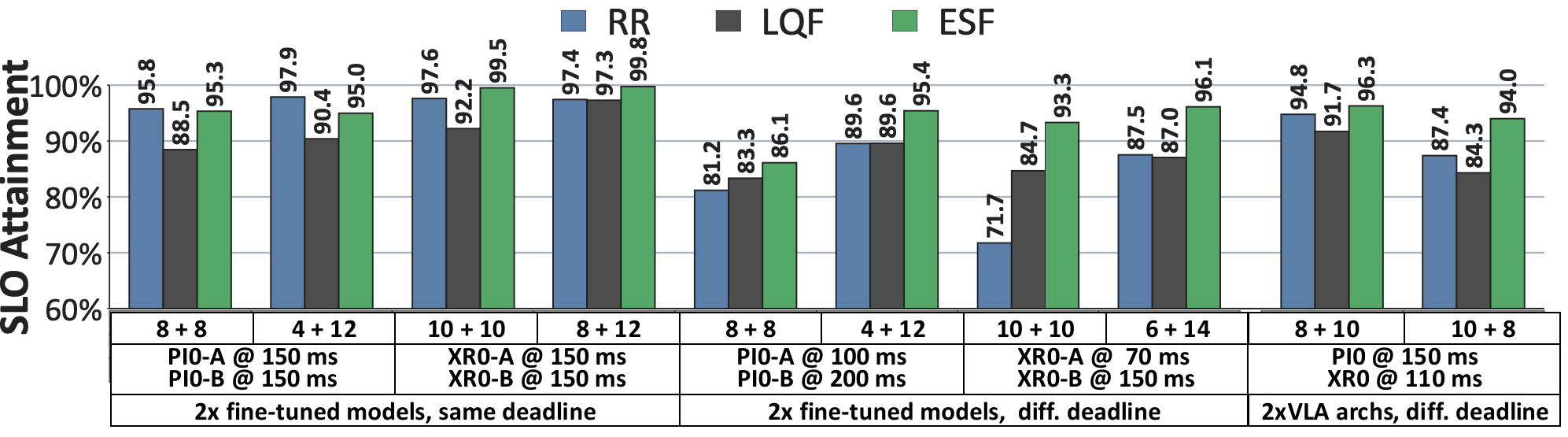}
    \vspace{-20pt}
    \caption{\SLOAT under three queue-selection policies, across multi-model co-location configurations with varying robot populations and SLOs per model.}
    \label{fig:scheduler-policy}
\end{figure}

\noindent\textbf{2) Analysis of Queue Selection Policies.} Fig.~\ref{fig:scheduler-policy} shows \SysName's \SLOAT under three queue-selection policies for choosing which per-model queue to serve next (§\ref{sec:mechanism-3}): (i) \textit{round-robin} (\textbf{RR}) cycles through each model's queue in turn; (ii) \textit{largest-queue-first} (\textbf{LQF}) always serves the queue with the most pending requests; and (iii) \textit{earliest-SLO-first} (\textbf{ESF}) is \SysName's policy (§\ref{sec:mechanism-3}). 
We evaluate configurations of co-located models with equal or differing SLO deadlines, and with varying robot populations across co-located models.



We make two observations. First, \textit{RR} achieves better \SLOAT than \textit{LQF} when co-located models have the same architecture and SLOs. In this case, all model queues need to be prioritized equally, and \textit{RR} captures this objective more effectively than \textit{LQF}. Second, \textit{ESF} offers little advantage when co-located models have the same architecture and SLOs, but clearly outperforms the other policies once architectures or SLOs vary, because \textit{RR} and \textit{LQF} lack SLO-awarenes and cannot prioritize queues by the requests' deadlines. Across all configurations, \textit{ESF} achieves 1.06$\times$ and 1.07$\times$ higher average \SLOAT than \textit{RR} and \textit{LQF}, respectively, confirming its robustness as \SysName's default policy.


\vspace{-2pt}
\subsection{\textbf{Multi-GPU Model Serving \& Traffic Control}}

\begin{figure}[H]
    \centering
    \includegraphics[width=\linewidth]{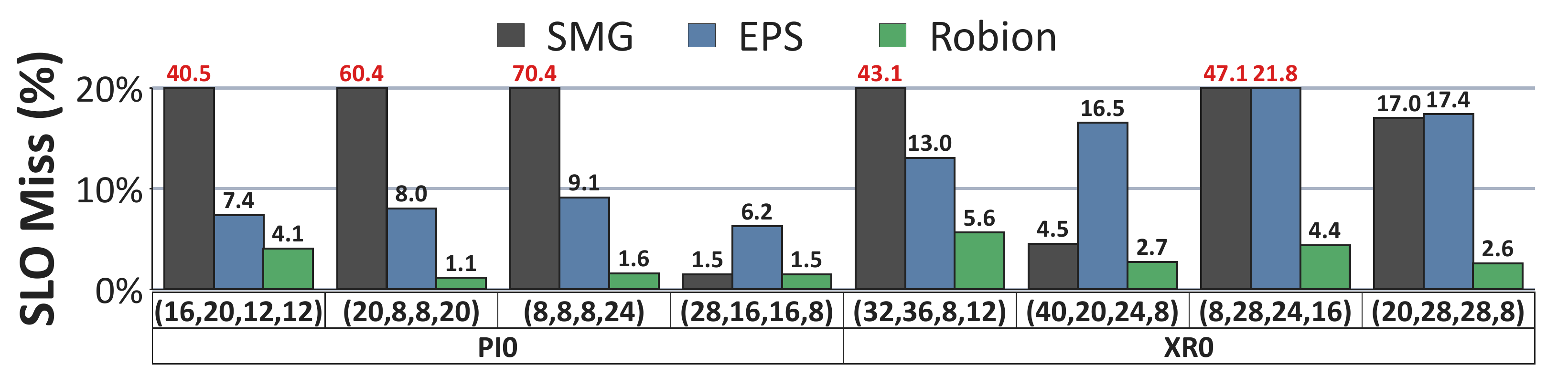}
    \vspace{-20pt}
    \caption{\SMR of three robot assignment strategies on a four-GPU server hosting four fine-tuned PI0 (left) and XR0 (right) variants, across varying robot populations.}
    \label{fig:allocation-frequency}
\end{figure}

\noindent\textbf{1) Benefit of \SysName's Traffic Adapter.} 
Fig.~\ref{fig:allocation-frequency} evaluates \SysName's traffic adapter (§\ref{sec:controller}) on a robot factory scenario with a 4-GPU server hosting 4 fine-tuned variants of PI0 (left) or XR0 (right), each GPU fits all 4 variants in its memory.
Each x-axis tuple denotes the robot population per model, with robot request frequencies of $1/2/2/4$ Hz, across the 4 models, respectively. We compare three strategies: (i) \textit{single-model-per-GPU} (\textbf{SMG}) (Fig.~\ref{fig:mechanism_traffic}a), where each model's robots send all requests to a single dedicated GPU that only hosts that model,
(ii) \textit{even-population-split} (\textbf{EPS})  (Fig.~\ref{fig:mechanism_traffic}b), where each model is replicated on every GPU and its robot population is split evenly across them, 
and (iii) \emph{\SysName} (Fig.~\ref{fig:mechanism_traffic}c) where each model is replicated on every GPU and its robot population is split across GPUs according to our integer program's solution. The per-model GPU limits (§~\ref{sec:controller}) across the 4 robot populations with varying request frequencies are $(27,16,16,9)$ for the PI0, and $(32,22,22,12)$ for the XR0.

We make three observations. 
First, when a model's robot population exceeds the GPU robot limit for that model, \textit{SMG} suffers substantial \SMR. For instance, at the 
$(8,8,8,24)$ population distribution for PI0, the fourth model's population $(24)$ far exceeds the GPU robot limit for that model $(9)$, and a single GPU cannot serve this excess traffic, resulting in a 70\% \SMR.
Second, when \emph{no} model's robot population significantly exceeds its GPU limit, e.g., the (28,16,16,8) population distribution for PI0, both \textit{SMG} and \SysName achieve very low \SMR, with \SysName's assignment converging to that of \textit{SMG}, whereas \textit{EPS} performs substantially worse, incurring 4.1$\times$ more \SMR in this example.
Third, \SysName's traffic adapter \underline{\emph{consistently}} and \underline{\emph{significantly}} outperforms the other two assignment strategies across all population distributions and both models, achieving on average 8.2$\times$ and 4.4$\times$ fewer SLO misses than \textit{SMG} and \textit{EPS}, respectively, combining the benefits of both approaches.

\begin{figure}[H]
    \centering
    \includegraphics[width=\linewidth]{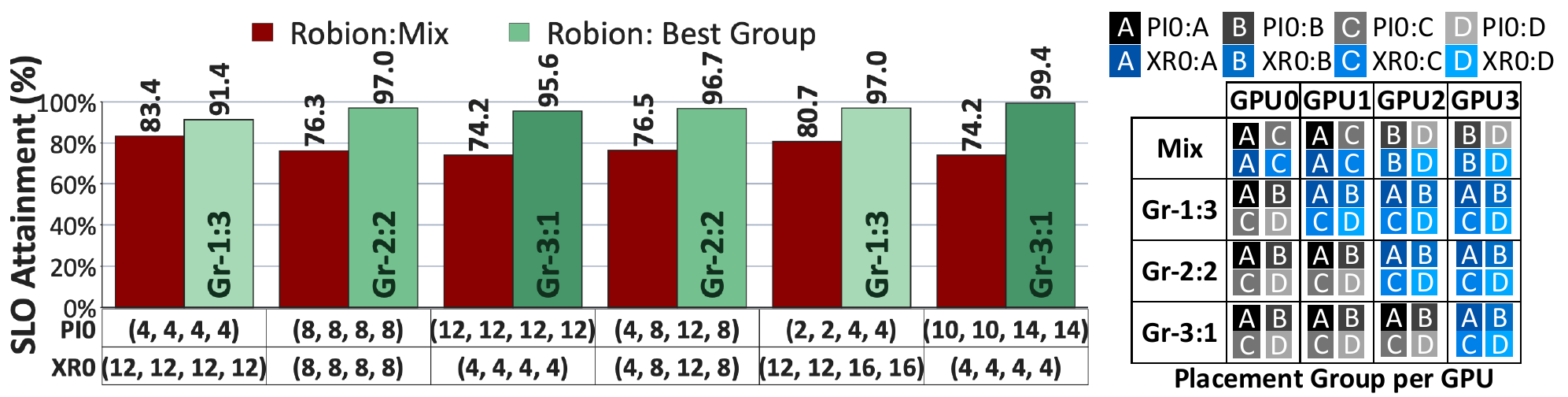}
    \vspace{-20pt}
    \caption{\SLOAT of \SysName under mixed vs. disjoint-group model placement for 8 PI0/XR0 variants on a 4-GPU server.}
    \vspace{-9pt}
    \label{fig:colocate-multiple-architectures}
\end{figure}

\noindent\textbf{2) Analysis of Various Model Placements.} 
Fig.~\ref{fig:colocate-multiple-architectures} shows \SysName \SLOAT for 8 fine-tuned VLA variants (4 from PI0, 4 from XR0), each with a different robot population, using a 4-GPU server under various model placements.
We compare two placement strategies. \emph{Mix} co-locates both architectures on every GPU: each GPU hosts 2 PI0 and 2 XR0 variants, with each variant replicated on 2 of the 4 GPUs. The alternative dedicates disjoint GPU groups per architecture, varying the group sizes: (i) \textit{Gr-1:3} assigns 1 GPU to PI0 and 3 GPUs to XR0, (ii) \textit{Gr-2:2} assigns 2 GPUs to each architecture, and (iii) \textit{Gr-3:1} assigns 3 GPUs to PI0 and 1 GPU to XR0, replicating all variants of an architecture across its assigned GPUs. For each robot population distribution, we report the best-performing placement among (i)-(iii), denoted \emph{Best-Group}, against \emph{Mix}.



We draw three findings. First, the best-performing placement among \textit{Gr-1:3}, \textit{Gr-2:2}, and \textit{Gr-3:1} varies with the robot population distribution across models. 
Second, \emph{Mix} consistently achieves lower \SLOAT than that of \emph{Best-Group} placement, since differing stage runtimes across co-located architectures introduce waiting delays for the model with shorter stage runtimes under \SysName's lockstep execution. 
Third, \SysName scales to large robot AI factories with high \SLOAT: in this setup, it serves 8 different models, i.e., 8 manipulation tasks, for up to \Largescalerobots robots (last column) within 98\% \SLOAT on a single 4-GPU server. We conclude that \SysName can accommodate large-scale robot factories spanning diverse tasks while serving several tens of robots on one edge server.



\begin{figure}[t]
    \centering
    \includegraphics[width=\linewidth]{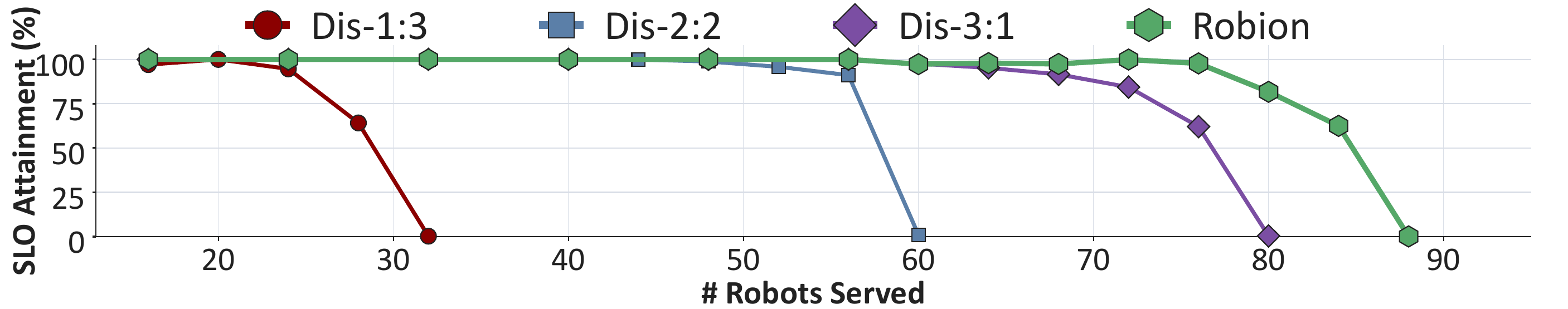}
    \vspace{-20pt}
    \caption{SLO attainment of \SysName versus inter-GPU stage disaggregation schemes for PI0 on a 4-GPU server.}
    \label{fig:disaggregation}
    \vspace{-10pt}
\end{figure}

\noindent\textbf{3) Analysis of Intra- vs. Inter-GPU Stage Disaggregation:} Fig.~\ref{fig:disaggregation} compares \SLOAT for PI0 on a 4-GPU server with \SysName, which disaggregates the VLM and ADiT stages within each GPU and replicates PI0 at all 4 GPUs, against three inter-GPU disaggregation schemes that instead dedicate separate GPUs to each stage, with no model replication: (i) \emph{Dis-1:3} uses 1 GPU for the VLM stage and 3 for the ADiT stage, (ii) \emph{Dis-2:2} uses 2 GPUs for each stage, and (iii) \emph{Dis-3:1} uses 3 GPUs for the VLM stage and 1 for the ADiT stage.


We make two observations. First, among the inter-GPU disaggregation schemes, \textit{Dis-1:3} performs worst, since running the memory-intensive ADiT stage on 3 GPUs leaves all of them underutilized (§\ref{sec:limitation_existing_systems}). Second, \SysName's intra-GPU disaggregation outperforms all inter-GPU schemes, serving 72 robots within 98\% \SLOAT versus 60 for the best-performing inter-GPU scheme, \textit{Dis-3:1}. This is because intra-GPU disaggregation replicates the entire model on all 4 GPUs, enabling independent request batches to run fully in parallel on all GPUs, while inter-GPU disaggregation replicates only a single stage across at most 3 GPUs.



\subsection{\SysName in an Enterprise Scenario on Cloud}

\begin{figure}[H]
    \vspace{-3pt}
    \centering
    \includegraphics[width=\linewidth]{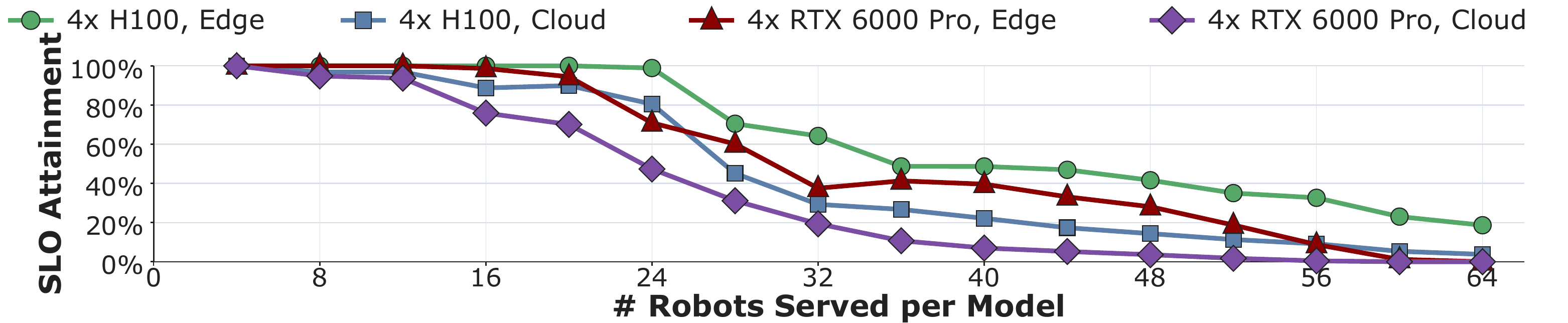}
    \vspace{-20pt}
    \caption{\SysName's \SLOAT serving 4 VLA models across different deployment settings and hardware platforms.} 
    \vspace{-8pt}
    \label{fig:cloud-comparison}
\end{figure}

\SysName can also be used for cloud deployment: robot factory owners could rent  cloud-hosted GPUs and use \SysName to serve multi-robot inference requests for their VLA models on remote cloud servers.
Fig.~\ref{fig:cloud-comparison} shows \SysName's \SLOAT as the number of robots served for each model increases, using 4 VLA model architectures (PI0, XR0, GR15, SVLA), each placed on a separate GPU. We evaluate four configurations, combining two server setups, 4$\times$RTX 6000 Pro and 4$\times$H100, with two deployment settings, edge (WiFi 7 network~\cite{Jiang2026HowFastCanIRunVLA}) and cloud (Fast Cloud network~\cite{Jiang2026HowFastCanIRunVLA}).

We draw two findings. First, cloud deployment achieves substantially lower \SLOAT than edge deployment due to higher network latencies: at 24 robots per-model, cloud deployment achieves 23.6\% and 18.4\% lower \SLOAT on the 4$\times$RTX 6000 Pro and 4$\times$H100 server, respectively. Edge servers, that host server-class GPUs while avoiding the cloud's network cost, are thus preferable. Second, on the edge deployment, the 4$\times$H100 configuration serves 24 robots within 98\% \SLOAT at about 2$\times$  the cost of the 4$\times$RTX 6000 Pro configuration, which serves 16 robots within 98\% \SLOAT.



\vspace{-3pt}
\section{Related Work}
\vspace{-1pt}
To our knowledge, this is the first work to (i) investigate the design considerations for efficient VLA serving on edge servers in robot factories, and (ii) propose a serving and management system that meets SLO requirements for heterogeneous multi-robot requests across multiple models on multiple GPUs. We briefly discuss prior work.

\subparagraph{\textbf{VLA Systems.}} 

ROSA~\cite{Jiang2026ROSA} addresses  model placement across a local cluster, Kairos~\cite{dai2026kairos} the task-level efficiency, and Armory~\cite{Bansal2026ActionChunkScheduling} the request scheduling to avoid robot starvation.
These works operate at the model placement, task-planning, or request scheduling, treating the VLA stages as a \emph{single} execution pipeline. None of them functions as a serving system capable of managing concurrent execution and GPU sharing across requests or models. \SysName  system that handles multiple concurrent requests across multiple VLA models, disaggregates VLA stages across GPU SMs, and meets SLOs. 

\subparagraph{\textbf{VLA Model \& Compilation Optimization.}} 
Prior works propose (i) action-chunking~\cite{Black2025pi0,Song2025PdVla,Zhao2023LearningFineGrained,Pertsch2025FastEfficient} to generate future action sequences, (ii) model compression via distillation~\cite{Huang2026RTVLA}, visual-token pruning~\cite{Jiang2025TheBetterYouLearn}, and quantization~\cite{Park2025SaliencyAwareQuantized,Williams2025LiteVla}, or (iii) caching~\cite{Li2026Oxygen, Su2026ExecutionStateCapsules, Xu2025VlaCache} and fused kernel optimizations~\cite{Ma2025RunningVLAsAtReal, Agarwal2026Flashrt}. These optimize the VLA model or its compiled version, while \SysName operates at the serving layer and can thus directly support these optimizations for further serving-level gains.


\subparagraph{\textbf{Serving Systems with Stage Disaggregation.}}

Prior systems~\cite{Yin2026vllmomni,sglang_omni_2026,Jha2026Mstar} disaggregate multi-stage models to enable independent scheduling. As we analyze in  §\ref{sec:limitation_existing_systems}, these works primarily target stage disaggregation on separate GPU devices and larger-scale models whose stages run for hundreds of milliseconds. In Fig. ~\ref{fig:single-model-single-gpu_rtx6000}, we quantitatively compare with vLLM-Omni~\cite{Yin2026vllmomni}, the most widely used among them, and show that \SysName significantly outperforms it. VLA models with millisecond-scale stages cannot tolerate these systems' design, which incurs high inter-stage communication costs.


\subparagraph{\textbf{DiT Serving Systems.}} 
A few works~\cite{Runyu2026Tetriserve,Ye2026GENSERVE,Fang2024xDiT,Li2025Katz,huang2025dditdynamicresourceallocation,qiang2026gfditschedulingparallelismdiffusion}  serve DiT-based text-to-image and text-to-video models. These models consist of a single heavy DiT stage that dominates inference runtime (>99\%), so these systems focus on multi-GPU parallelization of that single stage. VLA models are substantially smaller than DiT models and can fit on a single GPU, making cross-GPU parallelization unnecessary. VLAs also comprise two  stages with distinct compute and memory characteristics. \SysName serves them efficiently via intra-GPU stage disaggregation to enable independent stage scheduling.

\subparagraph{\textbf{GPU Sharing  \& ML Sharing Systems.}} 

GPU sharing mechanisms~\cite{Tanasic2014EnablingPreemptive,Park2015Chimera},  NVIDIA’s MIG~\cite{NVIDIA2024MIG}, MPS~\cite{NVIDIA2026MPS}, and Green Contexts~\cite{NVIDIA2026GreenContext} allocate GPU resources across processes. Building on them, ML sharing systems time-multiplex the entire device across models~\cite{Gujarati2020ServingDNNLike}, spatially partition the SMs to different models~\cite{Seungbeom2022ServingHeterogeneous,Fei2023igniter,Shubha2024USHER,Duan2024MuxServe, bullet, Zhang2025ImprovingGPUSharing}, or introduce a software scheduling layer between co-located kernels of different models and the native GPU scheduler~\cite{Strati2024Orion,Ng2023Paella,Tan2023GPUPool,Han2022REEF,Zhao2025Tally,lithos}.
This layer intercepts kernel launches from co-located models, adding a few microseconds of dispatch overhead per launch~\cite{Han2022REEF}. Large models (e.g., LLMs) amortize this overhead over kernel runtimes of tens to hundreds of microseconds. However, VLA models have substantially shorter kernel times, 8--10 $\mu s$, comparable to the dispatch overhead itself, making these systems ill-suited for co-locating VLA models on a GPU. \SysName's co-location mechanism  operates at stage granularity rather than kernel-level, eliminating per-kernel dispatch overheads.

\subparagraph{\textbf{LLM Serving Systems.}} 
LLM serving systems optimize KV cache management~\cite{Kwon2023vllm,Ye2025Flashinfer,Zheng2024Sglang}, efficiently batch variable-length requests~\cite{Sheng2024FairnessInServing,Wu2026FastServe,Sun2024Llumnix,ExeGPT}, and disaggregate the prefill and decode stages across GPUs~\cite{Zhong2024DistServe,Patel2024splitWise,Qin2025moonCake, bullet}. VLA models have a small KV cache, fixed-length requests, and fit on one GPU, eliminating the need for inter-GPU disaggregation. Thus, LLM serving systems would not benefit VLA models.

\vspace{-6pt}
\section{Conclusion}

We explore 
the design considerations necessary for efficient VLA serving in robot factories. Building on them, we design \SysName, an SLO-aware serving and management system for  multi-robot requests across multiple VLA models on multi-GPU local edge servers. 
For individual models, \SysName serves on average \Omnihigherload$\times$ and \Monhigherload$\times$ higher robot load within 98\% SLO attainment over vLLM-Omni, the most widely used multi-stage serving system, and Monolithic, which runs VLM and ADiT as a single pipeline, respectively. In a large-scale
experiment of serving 8 different models on a 4-GPU server, \SysName can serve up to \Largescalerobots robots within 98\% SLO attainment.
We hope our work encourages further research on system and architecture design for physical AI in robot factories.


\balance
\bibliographystyle{unsrtnat}
\bibliography{references}

@misc{Jiang2026HowFastCanIRunVLA,
      title={How Fast Can I Run My VLA? Demystifying VLA Inference Performance with VLA-Perf}, 
      author={Wenqi Jiang and Jason Clemons and Karu Sankaralingam and Christos Kozyrakis},
      year={2026},
      eprint={2602.18397},
      archivePrefix={arXiv},
      primaryClass={cs.RO},
      url={https://arxiv.org/abs/2602.18397}, 
}

@misc{yan2026agilevlafewshotindustrialpose,
      title={Agile-VLA: Few-Shot Industrial Pose Rectification via Implicit Affordance Anchoring}, 
      author={Teng Yan and Zhengyang Pei and Chengyu Shi and Yue Yu and Yikun Chen and Zilong Zhu and Zelin Fang and Kaile Guo and Zihang Wang and Peigen Tian and Bingzhuo Zhong},
      year={2026},
      eprint={2603.22899},
      archivePrefix={arXiv},
      primaryClass={cs.RO},
      url={https://arxiv.org/abs/2603.22899}, 
}

@misc{yu2026surveyefficientvisionlanguageactionmodels,
      title={A Survey on Efficient Vision-Language-Action Models}, 
      author={Zhaoshu Yu and Bo Wang and Pengpeng Zeng and Haonan Zhang and Ji Zhang and Zheng Wang and Lianli Gao and Jingkuan Song and Nicu Sebe and Heng Tao Shen},
      year={2026},
      eprint={2510.24795},
      archivePrefix={arXiv},
      primaryClass={cs.CV},
      url={https://arxiv.org/abs/2510.24795}, 
}

@techreport{pohland2026offload,
author = {Pohland, Sara and Foukas, Xenofon and Ananthanarayanan, Ganesh and Kolobov, Andrey and Mehrotra, Sanjeev and Radunovic, Bozidar and Verma, Ankit},
title = {Offload or Overload: A Platform Measurement Study of Mobile Robotic Manipulation Workloads},
institution = {Microsoft},
year = {2026},
month = {March},
url = {https://www.microsoft.com/en-us/research/publication/offload-or-overload-a-platform-measurement-study-of-mobile-robotic-manipulation-workloads/},
number = {MSR-TR-2026-14},
}

@inproceedings{
wang2026realtime,
title={Real-Time Robot Execution with Masked Action Chunking},
author={Haoxuan Wang and Gengyu Zhang and Yan Yan and Yuzhang Shang and Ramana Rao Kompella and Gaowen Liu},
booktitle={Proceedings of the International Conference on Learning Representations (ICLR)},
year={2026},
}

@INPROCEEDINGS{Park2025SaliencyAwareQuantized,
  author={Park, Seongmin and Kim, Hyungmin and Kim, Sangwoo and Jeon, Wonseok and Yang, Juyoung and Jeon, Byeongwook and Oh, Yoonseon and Choi, Jungwook},
  booktitle={2025 IEEE/CVF International Conference on Computer Vision (ICCV)}, 
  title={Saliency-Aware Quantized Imitation Learning for Efficient Robotic Control}, 
  year={2025},
  volume={},
  number={},
  pages={13140-13150},
  doi={10.1109/ICCV51701.2025.01221}}

@misc{Ma2025RunningVLAsAtReal,
      title={Running VLAs at Real-time Speed}, 
      author={Yunchao Ma and Yizhuang Zhou and Yunhuan Yang and Tiancai Wang and Haoqiang Fan},
      year={2025},
      eprint={2510.26742},
      archivePrefix={arXiv},
      primaryClass={cs.RO},
      url={https://arxiv.org/abs/2510.26742}, 
}

@misc{Zhao2025VLARAIL,
      title={VLA-RAIL: A Real-Time Asynchronous Inference Linker for VLA Models and Robots}, 
      author={Yongsheng Zhao and Lei Zhao and Baoping Cheng and Gongxin Yao and Xuanzhang Wen and Han Gao},
      year={2025},
      eprint={2512.24673},
      archivePrefix={arXiv},
      primaryClass={cs.RO},
      url={https://arxiv.org/abs/2512.24673}, 
}

@misc{Agarwal2026Flashrt,
      title={FlashRT: Agent Harness for Guiding Agents to Deploy Real-Time Multimodal Applications}, 
      author={Krish Agarwal and Zhuoming Chen and Yanyuan Qin and Zhenyu Gu and Atri Rudra and Beidi Chen},
      year={2026},
      eprint={2607.18171},
      archivePrefix={arXiv},
      primaryClass={cs.LG},
      url={https://arxiv.org/abs/2607.18171}, 
}

@inproceedings{Xu2025VlaCache,
    title={{VLA}-Cache: Efficient Vision-Language-Action Manipulation via Adaptive Token Caching},
    author={Siyu Xu and Yunke Wang and Chenghao Xia and Dihao Zhu and Tao Huang and Chang Xu},
    booktitle={The Thirty-ninth Annual Conference on Neural Information Processing Systems},
    year={2025},
    url={https://openreview.net/forum?id=QZYZ0Xm58q}
}

@article{Song2025PdVla,
  title={PD-VLA: Accelerating Vision-Language-Action Model Integrated with Action Chunking via Parallel Decoding},
  author={Wenxuan Song and Jiayi Chen and Pengxiang Ding and Han Zhao and Wei Zhao and Zhide Zhong and Zongyuan Ge and Jun Ma and Haoang Li},
  journal={2025 IEEE/RSJ International Conference on Intelligent Robots and Systems (IROS)},
  year={2025},
  pages={13162-13169},
  url={https://api.semanticscholar.org/CorpusID:276770282}
}

@misc{Huang2026RTVLA,
      title={RT-VLA: Real-Time Vision-Language-Action Models via Knowledge Distillation}, 
      author={Xiangyu Huang and Zhenlin Hua and Han Zhou and Shounak Sural and Ragunathan Rajkumar},
      year={2026},
      eprint={2606.14010},
      archivePrefix={arXiv},
      primaryClass={cs.CV},
      url={https://arxiv.org/abs/2606.14010}, 
}

@misc{Jiang2025TheBetterYouLearn,
      title={The Better You Learn, The Smarter You Prune: Towards Efficient Vision-language-action Models via Differentiable Token Pruning}, 
      author={Titong Jiang and Xuefeng Jiang and Yuan Ma and Xin Wen and Bailin Li and Kun Zhan and Peng Jia and Yahui Liu and Sheng Sun and Xianpeng Lang},
      year={2025},
      eprint={2509.12594},
      archivePrefix={arXiv},
      primaryClass={cs.RO},
      url={https://arxiv.org/abs/2509.12594}, 
}

@misc{Williams2025LiteVla,
      title={Lite VLA: Efficient Vision-Language-Action Control on CPU-Bound Edge Robots}, 
      author={Justin Williams and Kishor Datta Gupta and Roy George and Mrinmoy Sarkar},
      year={2025},
      eprint={2511.05642},
      archivePrefix={arXiv},
      primaryClass={cs.RO},
      url={https://arxiv.org/abs/2511.05642}, 
}

@misc{lu2026fasterrethinkingrealtimeflow,
      title={FASTER: Rethinking Real-Time Flow VLAs}, 
      author={Yuxiang Lu and Zhe Liu and Xianzhe Fan and Zhenya Yang and Jinghua Hou and Junyi Li and Kaixin Ding and Hengshuang Zhao},
      year={2026},
      eprint={2603.19199},
      archivePrefix={arXiv},
      primaryClass={cs.RO},
      url={https://arxiv.org/abs/2603.19199}, 
}

@misc{Nvidia2025GROOT,
      title={GR00T N1: An Open Foundation Model for Generalist Humanoid Robots}, 
      author={NVIDIA and : and Johan Bjorck and Fernando Castañeda and Nikita Cherniadev and Xingye Da and Runyu Ding and Linxi "Jim" Fan and Yu Fang and Dieter Fox and Fengyuan Hu and Spencer Huang and Joel Jang and Zhenyu Jiang and Jan Kautz and Kaushil Kundalia and Lawrence Lao and Zhiqi Li and Zongyu Lin and Kevin Lin and Guilin Liu and Edith Llontop and Loic Magne and Ajay Mandlekar and Avnish Narayan and Soroush Nasiriany and Scott Reed and You Liang Tan and Guanzhi Wang and Zu Wang and Jing Wang and Qi Wang and Jiannan Xiang and Yuqi Xie and Yinzhen Xu and Zhenjia Xu and Seonghyeon Ye and Zhiding Yu and Ao Zhang and Hao Zhang and Yizhou Zhao and Ruijie Zheng and Yuke Zhu},
      year={2025},
      eprint={2503.14734},
      archivePrefix={arXiv},
      primaryClass={cs.RO},
      url={https://arxiv.org/abs/2503.14734}, 
}

@ARTICLE{11164279,
  author={Kawaharazuka, Kento and Oh, Jihoon and Yamada, Jun and Posner, Ingmar and Zhu, Yuke},
  journal={IEEE Access}, 
  title={Vision-Language-Action Models for Robotics: A Review Towards Real-World Applications}, 
  year={2025},
  volume={13},
  number={},
  pages={162467-162504},
  doi={10.1109/ACCESS.2025.3609980}}

@article{Ma_2026,
   title={A Survey on Vision--Language--Action Models for Embodied AI},
   volume={37},
   ISSN={2162-2388},
   url={http://dx.doi.org/10.1109/TNNLS.2025.3650584},
   DOI={10.1109/tnnls.2025.3650584},
   number={7},
   journal={IEEE Transactions on Neural Networks and Learning Systems},
   publisher={Institute of Electrical and Electronics Engineers (IEEE)},
   author={Ma, Yueen and Song, Zixing and Zhuang, Yuzheng and Hao, Jianye and King, Irwin},
   year={2026}}

@misc{shao2025largevlmbasedvisionlanguageactionmodels,
      title={Large VLM-based Vision-Language-Action Models for Robotic Manipulation: A Survey}, 
      author={Rui Shao and Wei Li and Lingsen Zhang and Renshan Zhang and Zhiyang Liu and Ran Chen and Liqiang Nie},
      year={2025},
      eprint={2508.13073},
      archivePrefix={arXiv},
      primaryClass={cs.RO},
      url={https://arxiv.org/abs/2508.13073}, 
}

@misc{vllm_omni,
  author  = {{vLLM Project}},
  title   = {{vLLM-Omni}: A Framework for Efficient Model Inference
             with Omni-Modality Models},
  url     = {https://github.com/vllm-project/vllm-omni},
  yaer = {2026} 
}

@software{sglang_omni_2026,
  author  = {{SGLang Project Contributors}},
  title   = {{SGLang-Omni}: A High-Performance Serving Framework for
             Audio Models and Unified Multimodal Models},
  year    = {2026},
  version = {0.1.3},
  url     = {https://github.com/sgl-project/sglang-omni},
}

@inproceedings{
    su2025seesaw,
    title={Seesaw: High-throughput {LLM} Inference via Model Re-sharding},
    author={Qidong Su and Wei Zhao and Xin Li and Muralidhar Andoorveedu and Chenhao Jiang and Zhanda Zhu and Kevin Song and Christina Giannoula and Gennady Pekhimenko},
    booktitle={Eighth Conference on Machine Learning and Systems},
    year={2025},
    url={https://openreview.net/forum?id=YDksTjY3YY}
}

@misc{qiang2026gfditschedulingparallelismdiffusion,
      title={GF-DiT: Scheduling Parallelism for Diffusion Transformer Serving}, 
      author={Xinwei Qiang and Yifan Hu and Shixuan Sun and Jing Yang and Han Zhao and Chen Chen and Yu Feng and Jingwen Leng and Minyi Guo},
      year={2026},
      eprint={2606.13501},
      archivePrefix={arXiv},
      primaryClass={cs.DC},
      url={https://arxiv.org/abs/2606.13501}, 
}

@misc{huang2025dditdynamicresourceallocation,
      title={DDiT: Dynamic Resource Allocation for Diffusion Transformer Model Serving}, 
      author={Heyang Huang and Cunchen Hu and Jiaqi Zhu and Ziyuan Gao and Liangliang Xu and Yizhou Shan and Yungang Bao and Sun Ninghui and Tianwei Zhang and Sa Wang},
      year={2025},
      eprint={2506.13497},
      archivePrefix={arXiv},
      primaryClass={cs.DC},
      url={https://arxiv.org/abs/2506.13497}, 
}

@misc{holmes2024deepspeedfastgenhighthroughputtextgeneration,
      title={DeepSpeed-FastGen: High-throughput Text Generation for LLMs via MII and DeepSpeed-Inference}, 
      author={Connor Holmes and Masahiro Tanaka and Michael Wyatt and Ammar Ahmad Awan and Jeff Rasley and Samyam Rajbhandari and Reza Yazdani Aminabadi and Heyang Qin and Arash Bakhtiari and Lev Kurilenko and Yuxiong He},
      year={2024},
      eprint={2401.08671},
      archivePrefix={arXiv},
      primaryClass={cs.PF},
      url={https://arxiv.org/abs/2401.08671}, 
}

@misc{fastwam,
      title={Fast-WAM: Do World Action Models Need Test-time Future Imagination?}, 
      author={Tianyuan Yuan and Zibin Dong and Yicheng Liu and Hang Zhao},
      year={2026},
      eprint={2603.16666},
      archivePrefix={arXiv},
      primaryClass={cs.CV},
      url={https://arxiv.org/abs/2603.16666}, 
}

@misc{alpamayo,
      title={Alpamayo-R1: Bridging Reasoning and Action Prediction for Generalizable Autonomous Driving in the Long Tail}, 
      author={NVIDIA and : and Yan Wang and Wenjie Luo and Junjie Bai and Yulong Cao and Tong Che and Ke Chen and Yuxiao Chen and Jenna Diamond and Yifan Ding and Wenhao Ding and Liang Feng and Greg Heinrich and Jack Huang and Peter Karkus and Boyi Li and Pinyi Li and Tsung-Yi Lin and Dongran Liu and Ming-Yu Liu and Langechuan Liu and Zhijian Liu and Jason Lu and Yunxiang Mao and Pavlo Molchanov and Lindsey Pavao and Zhenghao Peng and Mike Ranzinger and Ed Schmerling and Shida Shen and Yunfei Shi and Sarah Tariq and Ran Tian and Tilman Wekel and Xinshuo Weng and Tianjun Xiao and Eric Yang and Xiaodong Yang and Yurong You and Xiaohui Zeng and Wenyuan Zhang and Boris Ivanovic and Marco Pavone},
      year={2026},
      eprint={2511.00088},
      archivePrefix={arXiv},
      primaryClass={cs.RO},
      url={https://arxiv.org/abs/2511.00088}, 
}

@misc{dream0,
      title={World Action Models are Zero-shot Policies}, 
      author={Seonghyeon Ye and Yunhao Ge and Kaiyuan Zheng and Shenyuan Gao and Sihyun Yu and George Kurian and Suneel Indupuru and You Liang Tan and Chuning Zhu and Jiannan Xiang and Ayaan Malik and Kyungmin Lee and William Liang and Nadun Ranawaka and Jiasheng Gu and Yinzhen Xu and Guanzhi Wang and Fengyuan Hu and Avnish Narayan and Johan Bjorck and Jing Wang and Gwanghyun Kim and Dantong Niu and Ruijie Zheng and Yuqi Xie and Jimmy Wu and Qi Wang and Ryan Julian and Danfei Xu and Yilun Du and Yevgen Chebotar and Scott Reed and Jan Kautz and Yuke Zhu and Linxi "Jim" Fan and Joel Jang},
      year={2026},
      eprint={2602.15922},
      archivePrefix={arXiv},
      primaryClass={cs.RO},
      url={https://arxiv.org/abs/2602.15922}, 
}

@inproceedings {orca,
author = {Gyeong-In Yu and Joo Seong Jeong and Geon-Woo Kim and Soojeong Kim and Byung-Gon Chun},
title = {Orca: A Distributed Serving System for {Transformer-Based} Generative Models},
booktitle = {16th USENIX Symposium on Operating Systems Design and Implementation (OSDI 22)},
year = {2022},
}

@misc{chittyvenkata2025moeinferencebenchperformanceevaluationmixture,
      title={MoE-Inference-Bench: Performance Evaluation of Mixture of Expert Large Language and Vision Models}, 
      author={Krishna Teja Chitty-Venkata and Sylvia Howland and Golara Azar and Daria Soboleva and Natalia Vassilieva and Siddhisanket Raskar and Murali Emani and Venkatram Vishwanath},
      year={2025},
      eprint={2508.17467},
      archivePrefix={arXiv},
      primaryClass={cs.LG},
      url={https://arxiv.org/abs/2508.17467}, 
}

@misc{zhong2025surveyvisionlanguageactionmodelsaction,
      title={A Survey on Vision-Language-Action Models: An Action Tokenization Perspective}, 
      author={Yifan Zhong and Fengshuo Bai and Shaofei Cai and Xuchuan Huang and Zhang Chen and Xiaowei Zhang and Yuanfei Wang and Shaoyang Guo and Tianrui Guan and Ka Nam Lui and Zhiquan Qi and Yitao Liang and Yuanpei Chen and Yaodong Yang},
      year={2025},
      eprint={2507.01925},
      archivePrefix={arXiv},
      primaryClass={cs.RO},
      url={https://arxiv.org/abs/2507.01925}, 
}

@inproceedings{Black2025pi0,
      title={$\pi_0$: A Vision-Language-Action Flow Model for General Robot Control}, 
      author={Kevin Black and Noah Brown and Danny Driess and Adnan Esmail and Michael Equi and Chelsea Finn and Niccolo Fusai and Lachy Groom and Karol Hausman and Brian Ichter and Szymon Jakubczak and Tim Jones and Liyiming Ke and Sergey Levine and Adrian Li-Bell and Mohith Mothukuri and Suraj Nair and Karl Pertsch and Lucy Xiaoyang Shi and James Tanner and Quan Vuong and Anna Walling and Haohuan Wang and Ury Zhilinsky},
      booktitle={Robotics: Science and Systems (RSS)},
      year={2025},
      url={https://arxiv.org/abs/2410.24164}, 
}

@inproceedings{Black2025pi05,
    title={{$\pi_{0.5}$}: A Vision-Language-Action Model with Open-World Generalization},
    author={Kevin Black and Noah Brown and James Darpinian and Karan Dhabalia and Danny Driess and Adnan Esmail and Michael Robert Equi and Chelsea Finn and Niccolo Fusai and Manuel Y. Galliker and Dibya Ghosh and Lachy Groom and Karol Hausman and Brian Ichter and Szymon Jakubczak and Tim Jones and Liyiming Ke and Devin LeBlanc and Sergey Levine and Adrian Li-Bell and Mohith Mothukuri and Suraj Nair and Karl Pertsch and Allen Z. Ren and Lucy Xiaoyang Shi and Laura Smith and Jost Tobias Springenberg and Kyle Stachowicz and James Tanner and Quan Vuong and Homer Walke and Anna Walling and Haohuan Wang and Lili Yu and Ury Zhilinsky},
    booktitle={9th Annual Conference on Robot Learning},
    year={2025},
    url={https://openreview.net/forum?id=vlhoswksBO}
}

@inproceedings{Zhao2023LearningFineGrained, 
    AUTHOR    = {Tony Z. Zhao AND Vikash Kumar AND Sergey Levine AND Chelsea Finn}, 
    TITLE     = {{Learning Fine-Grained Bimanual Manipulation with Low-Cost Hardware}}, 
    BOOKTITLE = {Proceedings of Robotics: Science and Systems}, 
    YEAR      = {2023}, 
    ADDRESS   = {Daegu, Republic of Korea}, 
    MONTH     = {July}, 
    DOI       = {10.15607/RSS.2023.XIX.016} 
}

@misc{Pertsch2025FastEfficient,
      title={FAST: Efficient Action Tokenization for Vision-Language-Action Models}, 
      author={Karl Pertsch and Kyle Stachowicz and Brian Ichter and Danny Driess and Suraj Nair and Quan Vuong and Oier Mees and Chelsea Finn and Sergey Levine},
      year={2025},
      eprint={2501.09747},
      archivePrefix={arXiv},
      primaryClass={cs.RO},
      url={https://arxiv.org/abs/2501.09747}, 
}

@misc{Cai2026XiaomiRobotics0,
      title={Xiaomi-Robotics-0: An Open-Sourced Vision-Language-Action Model with Real-Time Execution}, 
      author={Rui Cai and Jun Guo and Xinze He and Piaopiao Jin and Jie Li and Bingxuan Lin and Futeng Liu and Wei Liu and Fei Ma and Kun Ma and Feng Qiu and Heng Qu and Yifei Su and Qiao Sun and Dong Wang and Donghao Wang and Yunhong Wang and Rujie Wu and Diyun Xiang and Yu Yang and Hangjun Ye and Yuan Zhang and Quanyun Zhou},
      year={2026},
      eprint={2602.12684},
      archivePrefix={arXiv},
      primaryClass={cs.RO},
      url={https://arxiv.org/abs/2602.12684}, 
}

@misc{XiaomiRoboticsTeam2026XiaomiRobotics1,
      title={Xiaomi-Robotics-1: Scaling Vision-Language-Action Models with over 100K Hours of Real-World Trajectories}, 
      author={Xiaomi Robotics Team and Jun Guo and Piaopiao Jin and Jason Li and Peiyan Li and Yingyan Li and Futeng Liu and Wanli Peng and Optimus Qin and Yifei Su and Nan Sun and Qiao Sun and Runze Suo and Heyun Wang and Yunhong Wang and Rujie Wu and Caoyu Xia and Lina Zhang and Jack Zhao and Guoliang Chen and Wenlong Chen and Xinze He and Bin Li and Qing Li and Zhuorong Li and Heng Qu and Wenxuan Song and Diyun Xiang and Yifan Xie and Peiran Xu and Hangjun Ye and Wen Ye and Han Zhao and Quanyun Zhou},
      year={2026},
      eprint={2607.15330},
      archivePrefix={arXiv},
      primaryClass={cs.RO},
      url={https://arxiv.org/abs/2607.15330}, 
}

@misc{Shukor2025Smolvla,
      title={SmolVLA: A Vision-Language-Action Model for Affordable and Efficient Robotics}, 
      author={Mustafa Shukor and Dana Aubakirova and Francesco Capuano and Pepijn Kooijmans and Steven Palma and Adil Zouitine and Michel Aractingi and Caroline Pascal and Martino Russi and Andres Marafioti and Simon Alibert and Matthieu Cord and Thomas Wolf and Remi Cadene},
      year={2025},
      eprint={2506.01844},
      archivePrefix={arXiv},
      primaryClass={cs.LG},
      url={https://arxiv.org/abs/2506.01844}, 
}

@inproceedings{huang2026ticvla,
      title={TIC-VLA: A Think-in-Control Vision-Language-Action Model for Robot Navigation in Dynamic Environments},
      author={Zhiyu Huang and Yun Zhang and Johnson Liu and Rui Song and Chen Tang and Jiaqi Ma},
      booktitle={Proceedings of the International Conference on Machine Learning (ICML)},
      year={2026}
    }

@manual{NVIDIA2026MPS,
  title        = {Multi-Process Service (MPS)},
  author       = {{NVIDIA Corporation}},
  year         = {2026},
  url          = {https://docs.nvidia.com/deploy/mps/index.html},
  note         = {Accessed: 2026-08-13}
}

@manual{NVIDIA2026GreenContext,
  title        = {CUDA C++ Programming Guide: Green Contexts},
  author       = {{NVIDIA Corporation}},
  year         = {2026},
  url          = {https://docs.nvidia.com/cuda/cuda-programming-guide/04-special-topics/green-contexts.html},
  note         = {Accessed: 2026-08-13}
}

@inproceedings{Ng2023Paella,
    author = {Ng, Kelvin K. W. and Demoulin, Henri Maxime and Liu, Vincent},
    title = {Paella: Low-latency Model Serving with Software-defined GPU Scheduling},
    year = {2023},
    isbn = {9798400702297},
    publisher = {Association for Computing Machinery},
    address = {New York, NY, USA},
    url = {https://doi.org/10.1145/3600006.3613163},
    doi = {10.1145/3600006.3613163},
    booktitle = {Proceedings of the 29th Symposium on Operating Systems Principles},
    pages = {595–610},
    numpages = {16},
    location = {Koblenz, Germany},
    series = {SOSP '23}
}

@inproceedings{Zhao2025Tally,
    author = {Zhao, Wei and Jayarajan, Anand and Pekhimenko, Gennady},
    title = {Tally: Non-Intrusive Performance Isolation for Concurrent Deep Learning Workloads},
    year = {2025},
    isbn = {9798400706981},
    publisher = {Association for Computing Machinery},
    address = {New York, NY, USA},
    url = {https://doi.org/10.1145/3669940.3707282},
    doi = {10.1145/3669940.3707282},
    booktitle = {Proceedings of the 30th ACM International Conference on Architectural Support for Programming Languages and Operating Systems, Volume 1},
    pages = {1052–1068},
    numpages = {17},
    location = {Rotterdam, Netherlands},
    series = {ASPLOS '25}
}

@inproceedings{Strati2024Orion,
    author = {Strati, Foteini and Ma, Xianzhe and Klimovic, Ana},
    title = {Orion: Interference-aware, Fine-grained GPU Sharing for ML Applications},
    year = {2024},
    isbn = {9798400704376},
    publisher = {Association for Computing Machinery},
    address = {New York, NY, USA},
    url = {https://doi.org/10.1145/3627703.3629578},
    doi = {10.1145/3627703.3629578},
    booktitle = {Proceedings of the Nineteenth European Conference on Computer Systems},
    pages = {1075–1092},
    numpages = {18},
    location = {Athens, Greece},
    series = {EuroSys '24}
}

@inproceedings{Shubha2024USHER,
    author = {Shubha, Sudipta Saha and Shen, Haiying and Iyer, Anand},
    title = {USHER: holistic interference avoidance for resource optimized ML inference},
    year = {2024},
    isbn = {978-1-939133-40-3},
    publisher = {USENIX Association},
    address = {USA},
    booktitle = {Proceedings of the 18th USENIX Conference on Operating Systems Design and Implementation},
    articleno = {51},
    numpages = {18},
    location = {Santa Clara, CA, USA},
    series = {OSDI'24}
}

@misc{hirose2026asyncvlaasynchronousvlafast,
      title={AsyncVLA: An Asynchronous VLA for Fast and Robust Navigation on the Edge}, 
      author={Noriaki Hirose and Catherine Glossop and Dhruv Shah and Sergey Levine},
      year={2026},
      eprint={2602.13476},
      archivePrefix={arXiv},
      primaryClass={cs.RO},
      url={https://arxiv.org/abs/2602.13476}, 
}

@inproceedings{Agrawal2024TamingThroughputLatency,
    author = {Agrawal, Amey and Kedia, Nitin and Panwar, Ashish and Mohan, Jayashree and Kwatra, Nipun and Gulavani, Bhargav S. and Tumanov, Alexey and Ramjee, Ramachandran},
    title = {Taming throughput-latency tradeoff in LLM inference with sarathi-serve},
    year = {2024},
    isbn = {978-1-939133-40-3},
    publisher = {USENIX Association},
    address = {USA},
    booktitle = {Proceedings of the 18th USENIX Conference on Operating Systems Design and Implementation},
    articleno = {7},
    numpages = {18},
    location = {Santa Clara, CA, USA},
    series = {OSDI'24}
}

@misc{guo2026actioncontrolnetlightweightdelayaware,
      title={Action ControlNet: A Lightweight Delay-Aware Adapter for Smooth Asynchronous Control in Vision-Language-Action Models}, 
      author={Tiecheng Guo and Meng Guo},
      year={2026},
      eprint={2606.25985},
      archivePrefix={arXiv},
      primaryClass={cs.RO},
      url={https://arxiv.org/abs/2606.25985}, 
}

@inproceedings{Zhang2025ImprovingGPUSharing,
  title={Improving {GPU} Sharing Performance through Adaptive Bubbleless Spatial-Temporal Sharing},
  author={Zhang, Shulai and Chen, Quan and Cui, Weihao and Zhao, Han and Xue, Chunyu and Zheng, Zhen and Lin, Wei and Guo, Minyi},
  booktitle={Proceedings of the Twentieth European Conference on Computer Systems (EuroSys '25)},
  year={2025},
  publisher={ACM},
  address={New York, NY, USA},
  doi={10.1145/3689031.3696070},
  url={https://doi.org/10.1145/3689031.3696070}
}

@inproceedings {Seungbeom2022ServingHeterogeneous,
    author = {Seungbeom Choi and Sunho Lee and Yeonjae Kim and Jongse Park and Youngjin Kwon and Jaehyuk Huh},
    title = {Serving Heterogeneous Machine Learning Models on {Multi-GPU} Servers with {Spatio-Temporal} Sharing},
    booktitle = {2022 USENIX Annual Technical Conference (USENIX ATC 22)},
    year = {2022},
    isbn = {978-1-939133-29-53},
    address = {Carlsbad, CA},
    pages = {199--216},
    url = {https://www.usenix.org/conference/atc22/presentation/choi-seungbeom},
    publisher = {USENIX Association},
    month = jul
}

@inproceedings{Zhong2024DistServe,
    author = {Zhong, Yinmin and Liu, Shengyu and Chen, Junda and Hu, Jianbo and Zhu, Yibo and Liu, Xuanzhe and Jin, Xin and Zhang, Hao},
    title = {DistServe: disaggregating prefill and decoding for goodput-optimized large language model serving},
    year = {2024},
    isbn = {978-1-939133-40-3},
    publisher = {USENIX Association},
    address = {USA},
    booktitle = {Proceedings of the 18th USENIX Conference on Operating Systems Design and Implementation},
    articleno = {11},
    numpages = {18},
    location = {Santa Clara, CA, USA},
    series = {OSDI'24}
}

@manual{nvidia_cuda_driver,
  author       = {{NVIDIA Corporation}},
  title        = {{CUDA Driver API}},
  url          = {https://docs.nvidia.com/cuda/cuda-driver-api/}
}

@misc{amdcontexts,
  author       = {{Advanced Micro Devices, Inc.}},
  title        = {Execution contexts},
  howpublished = {\url{https://rocm.docs.amd.com/projects/HIP/en/develop/how-to/hip_runtime_api/execution_context.html}}
}

@misc{cudagraph,
  author       = {Gray, Alan},
  title        = {Getting Started with {CUDA} Graphs},
  year         = {2019},
  month        = sep,
  howpublished = {\url{https://developer.nvidia.com/blog/cuda-graphs/}},
  note         = {NVIDIA Technical Blog}
}

@misc{cuda_guide,
  author       = {{NVIDIA}},
  title        = {{CUDA C++ Programming Guide}},
  year         = {n.d.},
  howpublished = {\url{https://docs.nvidia.com/cuda/cuda-c-programming-guide/index.html}}
}

@inproceedings{Zheng2024Sglang,
    title={{SGL}ang: Efficient Execution of Structured Language Model Programs},
    author={Lianmin Zheng and Liangsheng Yin and Zhiqiang Xie and Chuyue Sun and Jeff Huang and Cody Hao Yu and Shiyi Cao and Christos Kozyrakis and Ion Stoica and Joseph E. Gonzalez and Clark Barrett and Ying Sheng},
    booktitle={The Thirty-eighth Annual Conference on Neural Information Processing Systems},
    year={2024},
    url={https://openreview.net/forum?id=VqkAKQibpq}
}

@inproceedings{Sheng2024FairnessInServing,
    author = {Sheng, Ying and Cao, Shiyi and Li, Dacheng and Zhu, Banghua and Li, Zhuohan and Zhuo, Danyang and Gonzalez, Joseph E. and Stoica, Ion},
    title = {Fairness in serving large language models},
    year = {2024},
    isbn = {978-1-939133-40-3},
    publisher = {USENIX Association},
    address = {USA},
    booktitle = {Proceedings of the 18th USENIX Conference on Operating Systems Design and Implementation},
    articleno = {52},
    numpages = {24},
    location = {Santa Clara, CA, USA},
    series = {OSDI'24}
}

@inproceedings{Sun2024Llumnix,
    author = {Sun, Biao and Huang, Ziming and Zhao, Hanyu and Xiao, Wencong and Zhang, Xinyi and Li, Yong and Lin, Wei},
    title = {Llumnix: dynamic scheduling for large language model serving},
    year = {2024},
    isbn = {978-1-939133-40-3},
    publisher = {USENIX Association},
    address = {USA},
    booktitle = {Proceedings of the 18th USENIX Conference on Operating Systems Design and Implementation},
    articleno = {10},
    numpages = {19},
    location = {Santa Clara, CA, USA},
    series = {OSDI'24}
}

@inproceedings{Ye2025Flashinfer,
  title={FlashInfer: Efficient and Customizable Attention Engine for {LLM} Inference Serving},
  author={Zihao Ye and Lequn Chen and Ruihang Lai and Wuwei Lin and Yineng Zhang and Stephanie Wang and Tianqi Chen and Baris Kasikci and Vinod Grover and Arvind Krishnamurthy and Luis Ceze},
  booktitle={MLSys 2025},
  year={2025},
  url={https://openreview.net/forum?id=RXPofAsL8F}
}

@misc{Yin2026vllmomni,
      title={vLLM-Omni: Fully Disaggregated Serving for Any-to-Any Multimodal Models}, 
      author={Peiqi Yin and Jiangyun Zhu and Han Gao and Chenguang Zheng and Yongxiang Huang and Taichang Zhou and Ruirui Yang and Weizhi Liu and Weiqing Chen and Canlin Guo and Didan Deng and Zifeng Mo and Cong Wang and James Cheng and Roger Wang and Hongsheng Liu},
      year={2026},
      eprint={2602.02204},
      archivePrefix={arXiv},
      primaryClass={cs.DC},
      url={https://arxiv.org/abs/2602.02204}, 
}

@misc{dai2026kairos,
      title={Kairos: A Scalable Serving System for Physical AI}, 
      author={Yinwei Dai and Ganesh Ananthanarayanan and Landon Cox and Xenofon Foukas and Bozidar Radunovic and Ravi Netravali},
      year={2026},
      eprint={2605.11381},
      archivePrefix={arXiv},
      primaryClass={cs.RO},
      url={https://arxiv.org/abs/2605.11381}, 
}

@software{ortools,
  title = {CP-SAT},
  version = { v9.12 },
  author = {Laurent Perron and Frédéric Didier},
  organization = {Google},
  url = {https://developers.google.com/optimization/cp/cp_solver/},
  date = { 2025-02-17 }
}

@inproceedings{Bansal2026ActionChunkScheduling, 
    title={Action Chunk Scheduling for Batched Robot Policy Serving}, author={Rohan Bansal and David He and Nadun Ranawaka Arachchige and Zhenyang Chen and Soobum Kim and Kexin Rong and Danfei Xu}, booktitle={Lab-to-Production Workshop}, year={2026}, url={https://openreview.net/forum?id=hxXIbCNwX9} 
}

@inproceedings{lithos,
author = {Coppock, Patrick H. and Zhang, Brian and Solomon, Eliot H. and Kypriotis, Vasilis and Yang, Leon and Sharma, Bikash and Schatzberg, Dan and Mowry, Todd C. and Skarlatos, Dimitrios},
title = {LithOS: An Operating System for Efficient Machine Learning on GPUs},
year = {2025},
isbn = {9798400718700},
publisher = {Association for Computing Machinery},
address = {New York, NY, USA},
url = {https://doi.org/10.1145/3731569.3764818},
doi = {10.1145/3731569.3764818},
booktitle = {Proceedings of the ACM SIGOPS 31st Symposium on Operating Systems Principles},
pages = {1–17},
numpages = {17},
location = {Lotte Hotel World, Seoul, Republic of Korea},
series = {SOSP '25}
}

@inproceedings{bullet,
author = {Lin, Zejia and Xu, Hongxin and Chen, Guanyi and Chen, Zhiguang and Lu, Yutong and Zhang, Xianwei},
title = {Bullet: Boosting GPU Utilization for LLM Serving via Dynamic Spatial-Temporal Orchestration},
year = {2026},
isbn = {9798400723599},
publisher = {Association for Computing Machinery},
address = {New York, NY, USA},
url = {https://doi.org/10.1145/3779212.3790135},
doi = {10.1145/3779212.3790135},
booktitle = {Proceedings of the 31st ACM International Conference on Architectural Support for Programming Languages and Operating Systems, Volume 2},
pages = {290–306},
numpages = {17},
location = {USA},
series = {ASPLOS '26}
}

@article{Qin2025moonCake,
author = {Qin, Ruoyu and Li, Zheming and He, Weiran and Cui, Jialei and Tang, Heyi and Ren, Feng and Ma, Teng and Cai, Shangming and Zhang, Yineng and Zhang, Mingxing and Wu, Yongwei and Zheng, Weimin and Xu, Xinran},
title = {Mooncake: A KVCache-centric Disaggregated Architecture for LLM Serving},
year = {2025},
publisher = {Association for Computing Machinery},
address = {New York, NY, USA},
issn = {1553-3077},
url = {https://doi.org/10.1145/3773772},
doi = {10.1145/3773772},
journal = {ACM Trans. Storage},
month = nov
}

@INPROCEEDINGS{Patel2024splitWise,
  author={Patel, Pratyush and Choukse, Esha and Zhang, Chaojie and Shah, Aashaka and Goiri, Íñigo and Maleki, Saeed and Bianchini, Ricardo},
  booktitle={2024 ACM/IEEE 51st Annual International Symposium on Computer Architecture (ISCA)}, 
  title={Splitwise: Efficient Generative LLM Inference Using Phase Splitting}, 
  year={2024},
  volume={},
  number={},
  pages={118-132},
  doi={10.1109/ISCA59077.2024.00019}}

@inproceedings{ExeGPT,
author = {Oh, Hyungjun and Kim, Kihong and Kim, Jaemin and Kim, Sungkyun and Lee, Junyeol and Chang, Du-seong and Seo, Jiwon},
title = {ExeGPT: Constraint-Aware Resource Scheduling for LLM Inference},
year = {2024},
isbn = {9798400703850},
publisher = {Association for Computing Machinery},
address = {New York, NY, USA},
url = {https://doi.org/10.1145/3620665.3640383},
doi = {10.1145/3620665.3640383},
booktitle = {Proceedings of the 29th ACM International Conference on Architectural Support for Programming Languages and Operating Systems, Volume 2},
pages = {369–384},
numpages = {16},
location = {La Jolla, CA, USA},
series = {ASPLOS '24}
}

@misc{Li2026Oxygen,
      title={OxyGen: Unified KV Cache Management for VLA Inference under Multi-Task Parallelism}, 
      author={Xiangyu Li and Huaizhi Tang and Xin Ding and Weijun Wang and Ting Cao and Yunxin Liu},
      year={2026},
      eprint={2603.14371},
      archivePrefix={arXiv},
      primaryClass={cs.RO},
      url={https://arxiv.org/abs/2603.14371}, 
}

@misc{Jiang2026ROSA,
      title={ROSA: A Robotics Foundation Model Serving System for Robot Factories}, 
      author={Wenqi Jiang and Jason Clemons and Rowland O'Flaherty and Hugo Hadfield and Alperen Degirmenci and Shuran Song and Yashraj Narang and Christos Kozyrakis},
      year={2026},
      eprint={2607.01088},
      archivePrefix={arXiv},
      primaryClass={cs.RO},
      url={https://arxiv.org/abs/2607.01088}, 
}

@inproceedings{Runyu2026Tetriserve,
author = {Lu, Runyu and He, Shiqi and Tan, Wenxuan and Li, Shenggui and Wu, Ruofan and Ma, Jeff J. and Chen, Ang and Chowdhury, Mosharaf},
title = {TetriServe: Efficiently Serving Mixed DiT Workloads},
year = {2026},
isbn = {9798400723599},
publisher = {Association for Computing Machinery},
address = {New York, NY, USA},
url = {https://doi.org/10.1145/3779212.3790233},
doi = {10.1145/3779212.3790233},
booktitle = {Proceedings of the 31st ACM International Conference on Architectural Support for Programming Languages and Operating Systems, Volume 2},
pages = {1982–1997},
numpages = {16},
location = {USA},
series = {ASPLOS '26}
}

@misc{Su2026ExecutionStateCapsules,
      title={Execution-State Capsules: Graph-Bound Execution-State Checkpoint and Restore for Low-Latency, Small-Batch, On-Device Physical-AI Serving}, 
      author={Liang Su},
      year={2026},
      eprint={2606.20537},
      archivePrefix={arXiv},
      primaryClass={cs.LG},
      url={https://arxiv.org/abs/2606.20537}, 
}

@inproceedings {Li2025Katz,
author = {Suyi Li and Lingyun Yang and Xiaoxiao Jiang and Hanfeng Lu and Dakai An and Zhipeng Di and Weiyi Lu and Jiawei Chen and Kan Liu and Yinghao Yu and Tao Lan and Guodong Yang and Lin Qu and Liping Zhang and Wei Wang},
title = {Katz: Efficient Workflow Serving for Diffusion Models with Many Adapters},
booktitle = {2025 USENIX Annual Technical Conference (USENIX ATC 25)},
year = {2025},
isbn = {978-1-939133-48-9},
address = {Boston, MA},
pages = {1037--1052},
url = {https://www.usenix.org/conference/atc25/presentation/li-suyi-katz},
publisher = {USENIX Association},
month = jul
}

@misc{Fang2024xDiT,
      title={xDiT: an Inference Engine for Diffusion Transformers (DiTs) with Massive Parallelism}, 
      author={Jiarui Fang and Jinzhe Pan and Xibo Sun and Aoyu Li and Jiannan Wang},
      year={2024},
      eprint={2411.01738},
      archivePrefix={arXiv},
      primaryClass={cs.DC},
      url={https://arxiv.org/abs/2411.01738}, 
}

@misc{Deng2025BAGEL,
      title={Emerging Properties in Unified Multimodal Pretraining}, 
      author={Chaorui Deng and Deyao Zhu and Kunchang Li and Chenhui Gou and Feng Li and Zeyu Wang and Shu Zhong and Weihao Yu and Xiaonan Nie and Ziang Song and Guang Shi and Haoqi Fan},
      year={2025},
      eprint={2505.14683},
      archivePrefix={arXiv},
      primaryClass={cs.CV},
      url={https://arxiv.org/abs/2505.14683}, 
}

@misc{Xu2025Qwen-Omni,
      title={Qwen2.5-Omni Technical Report}, 
      author={Jin Xu and Zhifang Guo and Jinzheng He and Hangrui Hu and Ting He and Shuai Bai and Keqin Chen and Jialin Wang and Yang Fan and Kai Dang and Bin Zhang and Xiong Wang and Yunfei Chu and Junyang Lin},
      year={2025},
      eprint={2503.20215},
      archivePrefix={arXiv},
      primaryClass={cs.CL},
      url={https://arxiv.org/abs/2503.20215}, 
}

@misc{zai2026glmimage,
  author       = {{Z.ai}},
  title        = {{GLM-Image}: Auto-regressive for Dense-knowledge
                  and High-fidelity Image Generation},
  year         = {2026},
  month        = jan,
  howpublished = {\url{https://huggingface.co/zai-org/GLM-Image}},
  note         = {AI model and model card, accessed 2026-08-16}
}

@inproceedings{Duan2024MuxServe,
author = {Duan, Jiangfei and Lu, Runyu and Duanmu, Haojie and Li, Xiuhong and Zhang, Xingcheng and Lin, Dahua and Stoica, Ion and Zhang, Hao},
title = {MuxServe: flexible spatial-temporal multiplexing for multiple LLM serving},
year = {2024},
publisher = {JMLR.org},
booktitle = {Proceedings of the 41st International Conference on Machine Learning},
articleno = {473},
numpages = {13},
location = {Vienna, Austria},
series = {ICML'24}
}

@inproceedings {Gujarati2020ServingDNNLike,
author = {Arpan Gujarati and Reza Karimi and Safya Alzayat and Wei Hao and Antoine Kaufmann and Ymir Vigfusson and Jonathan Mace},
title = {Serving {DNNs} like Clockwork: Performance Predictability from the Bottom Up},
booktitle = {14th USENIX Symposium on Operating Systems Design and Implementation (OSDI 20)},
year = {2020},
isbn = {978-1-939133-19-9},
pages = {443--462},
url = {https://www.usenix.org/conference/osdi20/presentation/gujarati},
publisher = {USENIX Association},
month = nov
}

@ARTICLE{Fei2023igniter,
  author={Xu, Fei and Xu, Jianian and Chen, Jiabin and Chen, Li and Shang, Ruitao and Zhou, Zhi and Liu, Fangming},
  journal={IEEE Transactions on Parallel and Distributed Systems}, 
  title={iGniter: Interference-Aware GPU Resource Provisioning for Predictable DNN Inference in the Cloud}, 
  year={2023},
  volume={34},
  number={3},
  pages={812-827},
  doi={10.1109/TPDS.2022.3232715}}

@inproceedings {Han2022REEF,
author = {Mingcong Han and Hanze Zhang and Rong Chen and Haibo Chen},
title = {Microsecond-scale Preemption for Concurrent {GPU-accelerated} {DNN} Inferences},
booktitle = {16th USENIX Symposium on Operating Systems Design and Implementation (OSDI 22)},
year = {2022},
isbn = {978-1-939133-28-1},
address = {Carlsbad, CA},
pages = {539--558},
url = {https://www.usenix.org/conference/osdi22/presentation/han},
publisher = {USENIX Association},
month = jul
}

@misc{Jha2026Mstar,
      title={M*: A Modular, Extensible, Serving System for Multimodal Models}, 
      author={Atindra Jha and Naomi Sagan and Keisuke Kamahori and Irmak Sivgin and Rohan Sanda and Steven Gao and Mark Horowitz and Luke Zettlemoyer and Olivia Hsu and Jure Leskovec and Baris Kasikci and Stephanie Wang},
      year={2026},
      eprint={2606.12688},
      archivePrefix={arXiv},
      primaryClass={cs.LG},
      url={https://arxiv.org/abs/2606.12688}, 
}

@inproceedings{Kwon2023vllm,
author = {Kwon, Woosuk and Li, Zhuohan and Zhuang, Siyuan and Sheng, Ying and Zheng, Lianmin and Yu, Cody Hao and Gonzalez, Joseph and Zhang, Hao and Stoica, Ion},
title = {Efficient Memory Management for Large Language Model Serving with PagedAttention},
year = {2023},
doi = {10.1145/3600006.3613165},
booktitle = {Proceedings of the 29th Symposium on Operating Systems Principles},
}

@inproceedings {Wu2026FastServe,
author = {Bingyang Wu and Yinmin Zhong and Zili Zhang and Shengyu Liu and Fangyue Liu and Yuanhang Sun and Gang Huang and Xuanzhe Liu and Xin Jin},
title = {{FastServe}: {Iteration-Level} Preemptive Scheduling for Large Language Model Inference},
booktitle = {23rd USENIX Symposium on Networked Systems Design and Implementation (NSDI 26)},
year = {2026},
isbn = {978-1-939133-54-0},
address = {Renton, WA},
pages = {57--74},
url = {https://www.usenix.org/conference/nsdi26/presentation/wu-bingyang},
publisher = {USENIX Association},
month = may
}

@misc{Ye2026GENSERVE,
      title={GENSERVE: Efficient Co-Serving of Heterogeneous Diffusion Model Workloads}, 
      author={Fanjiang Ye and Zhangke Li and Xinrui Zhong and Ethan Ma and Russell Chen and Kaijian Wang and Jingwei Zuo and Desen Sun and Ye Cao and Triston Cao and Myungjin Lee and Arvind Krishnamurthy and Yuke Wang},
      year={2026},
      eprint={2604.04335},
      archivePrefix={arXiv},
      primaryClass={cs.DC},
      url={https://arxiv.org/abs/2604.04335}, 
}

@inproceedings{Tan2023GPUPool,
author = {Tan, Xiaodan Serina and Golikov, Pavel and Vijaykumar, Nandita and Pekhimenko, Gennady},
title = {GPUPool: A Holistic Approach to Fine-Grained GPU Sharing in the Cloud},
year = {2023},
isbn = {9781450398688},
publisher = {Association for Computing Machinery},
address = {New York, NY, USA},
url = {https://doi.org/10.1145/3559009.3569650},
doi = {10.1145/3559009.3569650},
booktitle = {Proceedings of the International Conference on Parallel Architectures and Compilation Techniques},
pages = {317–332},
numpages = {16},
location = {Chicago, Illinois},
series = {PACT '22}
}

@inproceedings{Tanasic2014EnablingPreemptive,
author = {Tanasic, Ivan and Gelado, Isaac and Cabezas, Javier and Ramirez, Alex and Navarro, Nacho and Valero, Mateo},
title = {Enabling preemptive multiprogramming on GPUs},
year = {2014},
isbn = {9781479943944},
publisher = {IEEE Press},
booktitle = {Proceeding of the 41st Annual International Symposium on Computer Architecuture},
pages = {193–204},
numpages = {12},
location = {Minneapolis, Minnesota, USA},
series = {ISCA '14}
}

@misc{NVIDIA2024MIG,
  author       = {{NVIDIA Corporation}},
  title        = {NVIDIA Multi-Instance GPU (MIG) User Guide},
  year         = {2024},
  howpublished = {Available online},
  url          = {https://docs.nvidia.com/datacenter/tesla/mig-user-guide/latest/index.html}
}

@inproceedings{Park2015Chimera,
author = {Park, Jason Jong Kyu and Park, Yongjun and Mahlke, Scott},
title = {Chimera: Collaborative Preemption for Multitasking on a Shared GPU},
year = {2015},
isbn = {9781450328357},
publisher = {Association for Computing Machinery},
address = {New York, NY, USA},
url = {https://doi.org/10.1145/2694344.2694346},
doi = {10.1145/2694344.2694346},
booktitle = {Proceedings of the Twentieth International Conference on Architectural Support for Programming Languages and Operating Systems},
pages = {593–606},
numpages = {14},
location = {Istanbul, Turkey},
series = {ASPLOS '15}
}

@inproceedings{Liu2023LIBERO,
author = {Liu, Bo and Zhu, Yifeng and Gao, Chongkai and Feng, Yihao and Liu, Qiang and Zhu, Yuke and Stone, Peter},
title = {LIBERO: benchmarking knowledge transfer for lifelong robot learning},
year = {2023},
publisher = {Curran Associates Inc.},
address = {Red Hook, NY, USA},
booktitle = {Proceedings of the 37th International Conference on Neural Information Processing Systems},
articleno = {1939},
numpages = {16},
location = {New Orleans, LA, USA},
series = {NIPS '23}
}

@misc{bostondynamics2024stretchdhl,
  author       = {{Boston Dynamics}},
  title        = {Stretch at {DHL}},
  howpublished = {\url{https://bostondynamics.com/case-studies/stretch-at-dhl/}},
  year         = {2024},
  note         = {Accessed 2026-09-01},
  organization = {Boston Dynamics}
}

@misc{agility2023amazon,
  author       = {{Agility Robotics}},
  title        = {Agility Robotics Broadens Relationship with {Amazon}},
  howpublished = {\url{https://www.agilityrobotics.com/content/agility-robotics-broadens-relationship-with-amazon}},
  year         = {2023},
  note         = {Accessed 2026-09-01},
  organization = {Agility Robotics}
}

@misc{figure2026bmw,
  author       = {{Figure AI}},
  title        = {F.03 Arrives at {BMW}},
  howpublished = {\url{https://www.figure.ai/news/f-03-at-bmw}},
  year         = {2026},
  month        = jun,
  note         = {Accessed 2026-09-01},
  organization = {Figure AI}
}

\end{document}